\documentclass[nofootinbib,prx,aps,superscriptaddress,notitlepage,reprint,preprintnumbers,showpacs,showkeys]{revtex4-2}
\usepackage{latexsym,graphicx,amssymb,amsmath,mathrsfs,color}
\usepackage{physics}
\usepackage[utf8]{inputenc}
\usepackage{setspace,bm}
\usepackage{float}
\usepackage{subcaption}
\usepackage{placeins}
\usepackage[normalem]{ulem}

\usepackage{bbold}

\usepackage{wrapfig}
\usepackage{tikz}
\usepackage{hyperref}
\usetikzlibrary{decorations.pathmorphing,arrows.meta, positioning, calc, backgrounds}
\tikzset{
	fermion/.style={thick},
	boson/.style={
		thick,
		decorate,
		decoration={snake, amplitude=1.5pt}
		}
	}

\DeclareSymbolFont{bbold}{U}{bbold}{m}{n}
\DeclareSymbolFontAlphabet{\mathbbold}{bbold}

\newcommand{\be}{\begin{equation}}      
\newcommand{\ee}{\end{equation}}      
\newcommand{\bea}{\begin{eqnarray}}      
\newcommand{\eea}{\end{eqnarray}}

\definecolor{cfpastelpink}{RGB}{255,220,230}

\makeatletter
\renewcommand\appendix{\par
\setcounter{section}{0}
\setcounter{subsection}{0}
\gdef\thesection{\appendixname\space\@Alph\c@section}}

\long\def\unmarkedfootnote#1{{\long\def\@makefntext##1{##1}\footnotetext{#1}}}
\makeatother

\begin{document} 
\title{Utilizing Neural Networks for Solving Functional\\ Renormalization Group Flow Equations}

\author{J. Gurz\'o}
\email{gur.jo04@gmail.com}
\affiliation{Institute of Physics and Astronomy, E\"otv\"os University, 1117 Budapest, Hungary}
\author{G. Fej\H{o}s}
\email{gergely.fejos@ttk.elte.hu}
\affiliation{Institute of Physics and Astronomy, E\"otv\"os University, 1117 Budapest, Hungary}
\affiliation{Physics Division, Brookhaven National Laboratory, Upton, New York 11973, USA}

\begin{abstract}
A novel physics-informed neural networks (PINNs) based approach for solving functional renormalization group flow equations is presented. We investigate the finite temperature phase transition in $O(N)$ models without relying on the large-$N$ approximation, which was used in earlier attempts with similar methods. We argue that these previous solutions cannot easily be generalized for more complex symmetries, and hence our approach can provide new insight. Our technique relies on a series of ansätze that combines the small- and large-field approximation schemes with a neural network, leading to a highly accurate approximation over the entire field range. We compare the PINNs approach with traditional grid-based Runge-Kutta results and show that the network can successfully learn the temperature, field, and scale dependencies of the effective potential in the local potential approximation.
\end{abstract}

\maketitle

\section{Introduction}

The functional renormalization group (FRG) is one of the most prominent nonperturbative functional techniques for studying a wide range of phenomena in strongly coupled systems. It is extensively used in, but not limited to, e.g., statistical mechanics, particle physics, and quantum gravity; see the following extensive reviews \cite{Gies2012,Delamotte2012,Dupuis2021,Eichhorn2022}. The main idea behind the FRG is similar to that of the textbook Wilsonian renormalization group, namely, integrating out fluctuations from small to large wavelengths, but formulated in terms of the effective action, whose scale dependence is governed by Wetterich's flow equation \cite{Wetterich1993,Morris1993}.

The flow equation is a complicated functional integro-differential equation that must be approximated for practical calculations. Even after truncation, the resulting flow equations are often difficult to solve numerically, which makes it important to develop new techniques and approximation schemes that facilitate their analysis and improve the accuracy and efficiency of numerical calculations. Recent promising approaches include the reformulation of FRG flow equations as diffusive Hamilton--Jacobi-type equations \cite{Koenigstein2025}, the application of methods from computational fluid dynamics, such as finite-volume \cite{Koenigstein2022,Zorbach2024,Szep2026} and discontinuous Galerkin schemes \cite{Grossi2023,Ihssen2024,Sattler2026}, and the use of physics-informed neural networks (PINNs) \cite{Yokota2024,Tan2026,Yokozaki2026}. In this study, we aim to contribute to the development of the latter approach via investigations in the $O(N)$ symmetric scalar field theory.

The $O(N)$ scalar model is one of the most extensively studied quantum field theories, both within and beyond the framework of the FRG \cite{Juttner2017,DePolsi2020,Tan2024}. Its relative simplicity, combined with its critical behavior, makes it an important testing ground for nonperturbative methods and approximation schemes. In particular, the $O(N)$ model provides a useful framework for studying phase transitions, critical phenomena, and the structure of renormalization group flows. A wide variety of FRG truncation schemes have been applied to this model, making it a natural benchmark for assessing the accuracy and efficiency of new techniques.

In our present study we aim to develop a physics-informed neural network approach for solving FRG flow equations in the local potential approximation (LPA) for the $O(N)$ model. A similar attempt has already been made in a PINNs approach by Tan et al.~\cite{Tan2026}. In this paper, the authors rely on the large-$N$ approximation and make use of the implicit analytic solution for the effective potential in the $N\to\infty$ limit as a baseline, while a neural network is trained to learn the finite-$N$ corrections. The advantage of this clever decomposition is that the primary numerical challenge, the gradual flattening of the effective potential toward convexity in the infrared (IR), is largely captured by the large-$N$ solution. Consequently, the neural network only needs to learn the residual flattening.

The aforementioned method, while successful for the $O(N)$ model, comes at the expense of generality, since such a large-$N$ expansion is not necessarily available for more complex multi component scalar models. Even if a similar approximation scheme could be constructed, it is not guaranteed that the resulting flow equations would have an analytic solution. Furthermore, the earlier PINNs solution also made use of finite difference approximations to obtain the derivatives of the potential. While this choice is perfectly reasonable and simplifies numerical implementation, it may reduce numerical accuracy as finite difference techniques tend to become increasingly unstable, especially for multi variable functions, such as effective potentials of theories with more complicated symmetries  and thus invariant tensors.

In this paper, we present an alternative PINNs approach for solving the $O(N)$ model that aims to overcome these numerical challenges through a structured ansatz that can be generalized to more complicated symmetries. Our broader goal is to develop a method that can be applied to problems in which the effective potential depends on multiple variables, which tend to make the solution of the corresponding flow equations on a grid increasingly costly from a numerical point of view.

The paper is organized as follows. In Sec. II, we define our model and briefly discuss how to derive an approximate flow equation for the effective potential in the local potential approximation of the FRG formalism. In Sec. III, we discuss numerical methods: first, we review the ordinary grid-method, then introduce our PINNs approach to solve the (approximate) flow equation. Sec. IV is devoted for the presentation of the numerical results, while the reader can find our conclusions in Sec. V.

\section{$O(N)$ model in the FRG approach}

The $O(N)$ model is a field theory of $N$ real scalars, $\phi=(\phi_1,\phi_2,...\phi_N)$, defined by the following (Euclidean) Lagrangian: ($i=0,...,d$)
\bea
{\cal L} = \frac12 \partial_i \phi^T \partial_i \phi+\frac12m^2 \phi^T \phi+\frac{g}{4!}(\phi^T \phi)^2.
\eea
The Wetterich equation for the scale dependent effective action, $\Gamma_k$, reads
\bea
\label{Eq:Wet}
\partial_k \Gamma_k = \frac12 \Tr \big[\partial_k R_k (\Gamma_k^{(2)}+R_k)^{-1}],
\eea
where $(\Gamma_k^{(2)})_{ij}=\delta^2 \Gamma_k/ \delta \phi_i \delta \phi_j$ is the second derivative of $\Gamma_k$, while $R_k$ represents the regulator, whose role is to suppress IR modes with momenta $q\lesssim k$. The trace operation needs to be taken both in the functional and in the matrix senses. In this paper we work with the local potential approximation (LPA), where $\Gamma_k$ is approximated as
\bea
\label{Eq:LPA}
\Gamma_k = \int_0^\beta d\tau \int d^dx \Big[\frac12 \partial_i \phi^T \partial_i \phi+V_k(\rho)\Big],
\eea
where $\rho=(\phi^T \phi)/2$ is the $O(N)$ quartic invariant. The $x$ integral is over the $d$ dimensional real space, while $\tau$ corresponds to imaginary time, where $\beta=1/T$ is the inverse temperature. Plugging (\ref{Eq:LPA}) into (\ref{Eq:Wet}) we arrive at a flow equation for the effective potential, $V_k$:
\bea
\label{Eq:LPAflow}
\partial_k V_k &=& \frac{T}{2}\sum_{\omega_n} \int \frac{d^dq}{(2\pi)^d}{\rm{tr}}\Big[\partial_k R_k(\omega_n,{\bf q})\nonumber\\
&&\Big(\omega_n^2+{\bf q}^2+R_k(\omega_n,{\bf q})+V_k''\Big)^{-1}\Big],
\eea
which is now written explicitly in Fourier space. Note that the trace is now taken in the matrix sense only, and the corresponding Matsubara frequencies are $\omega_n = 2\pi n T$ ($n\in \mathbb{Z}$).

Before proceeding, a comment on the regulator is in order. We wish to work with Litim's optimal regulator \cite{Litim2001}, but we first need to decide whether to regulate both the frequency and the momenta, or only the momenta. The first choice corresponds to
\bea
R_k(\omega_n,{\bf q})=(k^2-{\bf q}^2-\omega_n^2)\Theta(k^2-{\bf q}^2-\omega_n^2)\mathbb{1}_{N\times N},
\eea
while the second one is simply
\bea
\label{Eq:reg2}
R_k(\omega_n,{\bf q})=(k^2-{\bf q}^2)\Theta(k^2-{\bf q}^2)\mathbb{1}_{N\times N}.
\eea
In this paper, we use only the second regulator, but we wish to mention that we attempted to work with the first one as well. We found, however, that it makes the flow numerically very unstable, both when using the grid method and when using PINNs. The results obtained this way are very inaccurate numerically, and we, therefore, do not wish to show any of the corresponding results, but advise the reader against using this choice.

Using (\ref{Eq:reg2}), the momentum integral becomes trivial and (\ref{Eq:LPAflow}) turns into
\bea
\label{Eq:LPAONflow}
\partial_k V_k &\!\!=\!\!& k^{d+1}\Omega_d T\sum_{\omega_n} \Bigg[\frac{N-1}{\omega_n^2+k^2+V_k'}\nonumber\\
&&\hspace{2.2cm}+\frac{1}{\omega_n^2+k^2+V_k'+2\rho V_k''}\Bigg]\nonumber\\
&\!\!=\!\!&k^{d+1}\Omega_d \Bigg[(N-1)\frac{1 + 2 n_B(\sqrt{k^2+V_k'})}{\sqrt{k^2+V_k'}}\nonumber\\
&&\hspace{1.5cm}+\frac{1 + 2 n_B(\sqrt{k^2+V_k'+2\rho V_k''})}{\sqrt{k^2+V_k'+2\rho V_k''}}\Bigg],
\eea
where $n_B(E)=1/(\exp(E/T)-1)$ is the Bose-Einstein distribution, and $\Omega_d=\int d\Omega_d/(2\pi)^d$ comes from the angular intergral.

As usual in the FRG formalism, we initialize the flow equation at the ultraviolet (UV) cutoff, $k=\Lambda$, by setting $\Gamma_\Lambda = S = \int_\tau\int_x \mathcal{L}$, where $S$ is the classical action, and then integrate the flow down to $k\rightarrow 0$ to obtain the full quantum effective action in the IR.

\section{Numerical methods}

In this section, we overview the numerical methods used in our study. Since these methods employ different numerical representations of the solution to the flow equation, it turns out to be useful to characterize them in terms of the number of degrees of freedom (DOF) involved.

In the context of numerically representing a function, such as the solution of the flow equation, the number of DOF may be defined as the independent real parameters admitted by the representation, which one can vary to scan through the representable functions. For a traditional finite element method, DOFs are function values on a lattice, whereas in the context of an ansatz based solution, DOFs are the tunable ansatz parameters. Note that when one needs an infrastructure to represent any possible function on some bounded domain, the finite element representation is clearly the best choice. Crucially, this is rarely ever the case in real physical applications, since the governing equations usually restrict the class of admissible solutions via boundary conditions, asymptotic behavior, differentiability, etc. Consequently, a finite element method will utilize its DOFs worse than a well formulated ansatz, when representing such a function. This, therefore, means that a grid based method will need significantly more DOFs to represent a function of a real physical problem, than a well formulated ansatz. The importance of this observation is best illustrated when considering the LPA flow equation of a theory with such a symmetry group that admits $n$ invariants\footnote{That is, the effective potential becomes an $n$ variable function.}. For a finite element method this implies an $n$ dimensional lattice (or $n+2$ if we count the $k$ and the $T$ dependence as well) which results in a $\#\text{DOF} \sim L^{n}$ with $L$ being the number of lattice points along one direction. The exponential scaling with $n$ results in rapidly increasing computational time and memory usage even when the complexity of the represented function does not grow at this rate, which means that the shown exponential scaling is not justified. This growth and the numerical instabilities arising from finite difference techniques in higher dimensions force us to conclude that grid-based methods are not capable of handling the flow equations of such theories. Consequently, one needs to find better numerical representations, which is exactly what we aim for with PINNs.

Since a PINNs based solution of the FRG flow equations does not have well established best practices, we focus on the $O(N)$ theory in the LPA. The theory admits a single invariant, therefore, the traditional grid based method can be utilized for solving the flow equation for the effective potential. A sufficiently fine discretization reproduces results of the more robust Kurganov--Tadmor finite-volume scheme~\cite{Szep2026}, and thus provide a reliable basis of comparison for the PINNs results.

For the sake of concrete purposes, we choose to investigate the $N=4$, $d=3$ theory. In all of the numerics, dimensionful quantities will be measured in units of the UV cutoff, i.e., $\Lambda = 1$. Note that, all couplings in the UV are chosen such that the critical temperature falls in the physically reasonable energy region: $0 < T_c < 1$. One particular choice of parameters that admits such a critical temperature is $m^2_\Lambda = -0.95$, $g_\Lambda = 120$, which will be used in the forthcoming analyses.

\subsection{Grid-based method}
Traditionally, the solution of the flow equation for the $O(N)$ effective potential, dependent on the orthogonal invariant, $\rho$, is obtained by uniformly sampling the domain between some endpoints and thus discretizing the problem. Here we use 1000 points covering the $\rho\in[0, 2 \rho_{0, \Lambda}]$ region, where $\rho_{0, \Lambda} = - 3 m_\Lambda^2/g_\Lambda$ is the minimum of $V_{k=\Lambda}$.

Functions are discretized by representing them with the array of values that the function takes on the grid points. As the first step of solving the flow equation, we initialize this array using the UV potential as follows,
\begin{align}
    V^{(i)}_{k=\Lambda} = V_\Lambda(\rho_i) =  m_\Lambda^2 \rho_i+\frac{g_\Lambda}{6}\rho_i^2\ .
\end{align}
To evaluate the right-hand side (rhs) of the flow equation, we need to determine $V_k'$ and $V_k''$ at the grid points. We do this with the finite difference approximation using 4 and 5 points. Equating the rhs of the flow equation to $\partial_k V_k$ at the lattice points, we can update the $V^{(i)}_{k}$ array elements using the {\it Runge-Kutta 45} method with adaptive step size\footnote{We used $\texttt{atol}=10^{-10}$, $\texttt{rtol}=0.0$ and $\texttt{dk\_guess}=-10^{-3}$, following standard conventions of {\it Runge-Kutta 45}.} in $k$. One simply needs to repeat this process, gradually reaching the IR region.

Although achieving $k\rightarrow 0$ would be desired, it gets numerically very costly since at some point, $k$ cannot compensate for the negative derivative(s) in the denominators of the eigenmodes of the propagator, and the expression becomes numerically unstable. Note that, this only occurs in the broken phase and its physical reason is the gradual flattening of the potential towards convexity. The phenomenon forces the numerical solver to accept increasingly smaller step sizes, which prohibits the simulation from reaching $k=0$ within a finite number of steps. Consequently, one needs to soften the escape condition of the simulation: terminate when a given number of iterations have passed or numerical instability arises.

In the grid method, the temperature, $T$, is always considered a parameter, therefore, for each value of $T$, we have to solve the flow equation. These are independent runs, and their parallel execution is straightforward using a thread pool.\vspace{-0.15cm}

\subsection{PINNs method}
In case of the PINNs, the solution of the flow equation is approximated by an ansatz with some tunable parameters. The primary objectives are finding the actual form of the ansatz and then its optimal parameters. The idea is very similar to that of the variational method in finding the lowest eigenvalue of the Hamiltonian of a quantum mechanical system. For the latter, one typically postulates an ansatz ground state $\ket{0;\theta_i}$ with tunable ansatz parameters ($\theta_i$) and minimizes
\begin{align}
 f(\theta_i) = \frac{\bra{0;\theta_i} \hat{H} \ket{0;\theta_i}}{\braket{0;\theta_i}{0;\theta_i}}\geq E_0\ ,
\end{align}
where $\hat{H}$ is the Hamiltonian and $E_0$ is the ground state energy. Note that, contrary to the Schrödinger equation, the FRG flow equations are highly nonlinear partial differential equations (PDEs), and thus turning them into an optimization problem is less straightforward. 

Here we show how to turn the task of solving the $O(N)$ LPA flow equation into an optimization problem, while only a slight generalization is required for applying it to any other PDEs~\cite{PINNs_the_og1,PINNs_the_og2}. Take a generic ansatz for the solution of the flow equation, $v(\rho,T,k;\theta_i)$ with tunable ansatz parameters $\theta_i$. Assuming that this is an explicit differentiable function, one can analytically compute $\partial_k v$, $\partial_\rho v$, $\partial_\rho^2 v$ or any other derivative. Hence, the derivatives are calculated exactly instead of approximating them with finite difference techniques, circumventing the problem of possible numerical instabilities. The flow equation can be written formally as
\begin{align}
    \partial_k V_k(\rho;T) = f\big(V_k'(\rho;T), V_k''(\rho;T), \rho, T, k\big)\ ,
\end{align}
where $f$ is the rhs of (\ref{Eq:LPAONflow}) and from here onward, the temperature is also considered a variable. This allows us to define the local PDE residue for the ansatz at a point $(\rho, T, k)$ as follows:
\begin{align}
    R(\rho, T, k; \theta_i) &= f\left(\partial_\rho v(\rho, T, k; \theta_i), \partial_\rho^2 v(\rho, T, k; \theta_i), \rho, T, k\right)\nonumber\\
    &- \partial_k v(\rho, T, k; \theta_i)\ .
\end{align}
This quantifies locally how badly the the ansatz solves the flow equation, for a given parameter set. Combining the PDE residues at all points of interest one arrives at the following PDE loss:
\begin{align}
    L_{\text{PDE}}(\theta_i) = \left(\int_{\rho, T, k} R^2(\rho, T, k; \theta_i) P_{\text{PDE}}(\rho, T, k) \right)^{1/2}\ ,
\end{align}
where $P_{\text{PDE}}$ is a nonnegative weight function for which the integral is well defined. The integration limits are such that $T\in[T_{\text{min}}, T_{\text{max}}]$, $\rho\in[0, \infty)$ and $k\in[0,1]$. In the special case of a fixed temperature training one has $P_{\text{PDE}}(T, k, \rho) \propto \delta(T - T_0)$, where $T_0$ is the temperature we are interested in. Crucially, for all ansatz parameters, $L_{\text{PDE}}(\theta_i)\geq 0$ and $L_{\text{PDE}}(\theta_i) = 0$ holds exactly when the PDE residue vanishes at every point, where $P_{\text{PDE}}>0$. Assuming that the ansatz satisfies the UV boundary condition, the optimal $\theta_i^*$ parameter set is the one that minimizes $L_{\text{PDE}}$. This is how one can translate a PDE to a minimization problem, which is the foundation of the PINNs approach. In our work we evaluated the PDE loss with
\begin{align}
    P_{\text{PDE}}(\rho,T,k) = \frac{1}{N}\sum_{j=1}^N\delta(\rho - \rho_j)\delta(T - T_j)\delta(k - k_j)\ ,
\end{align}
which turns the integral into a sum over all the $(\rho_j, T_j k_j)$ triplets. We use $N=300000$ for such triplets, which were randomly generated in each iteration of the minimization process. We refer to these triplets as the training batch.

In the context of machine learning, the minimization of the loss function is done using gradient descent or more elaborate gradient descent based algorithms, e.g., Adaptive Moment Estimation (ADAM)~\cite{Adam}. While this provides a robust minimization technique, to ensure convergence, we also employed second order Limited-memory Broyden–Fletcher–Goldfarb–Shanno (L-BFGS) optimizer \cite{Lbfgs_original} at the end of training.

Now we need to discuss how neural networks can be used for ansatz generation. In general, neural networks are functions with arbitrary number of inputs, outputs, and tunable ansatz parameters. They are known to be a robust tool of function approximation, due to the universal approximation theorem~\cite{deep_learning}, which loosely states that by increasing the number of tunable ansatz parameters one can approximate any continuous function to arbitrary precision with a neural network. 

The simplest, so-called ``feed-forward network'' (FFN) works with activation layers, which we may think of as column vectors. Let us consider a number of $D$ layers. To obtain the $l+1^{\text{st}}$ layer, first a linear transformation is applied on the $l^{\text{th}}$ layer, then an activation function, $\sigma$, acts element-wise on the result, introducing some form of nonlinearity. Mathematically, this process can be expressed as
\begin{align}
    \label{Eq:layers}
    a^{(l+1)}_k = \sigma\left(b^{(l+1)}_k + \sum_m W_{k m}^{(l+1)}a^{(l)}_m\right)\ ,
\end{align}
where $W^{(l)}$ are some weight matrices, while the $b^{(l)}$ columns are called the bias vectors. Eq. (\ref{Eq:layers}) shows how we get the $l+1^{\text{st}}$ layer of activation values from the previous one, see also the illustration in Fig. \ref{fig:FFN_unit_trans}. By the composition of such transformations we obtain the output from the input,
\begin{align}
    \label{Eq:NN}
    \mathcal{N}\ :\ \boldsymbol{a}^{\text{IN}}\to\overbrace{\boldsymbol{a}^{(1)}\to \dots \to \boldsymbol{a}^{(D)}}^{\text{hidden layers}} \to \boldsymbol{a}^{\text{OUT}}\ .
\end{align}
In our work, we use the $\sigma=\tanh$ activation function. Consequently, we drop the activation function for the last step in the composition since it would constrain the output(s) into the interval (-1,1). The $W^{(l)}$ weight matrices and the $b^{(l)}$ bias vectors play the role of tunable ansatz parameters. Therefore, by increasing the depth (i.e., the number of hidden layers, $D$) or the hidden size (i.e., the number of activation values, also called neurons on a hidden layer, HS) of the neural network one can increase the number of tunable ansatz parameters and thus the best achievable accuracy of representation of the function in question.

For the sake of simplicity, and on the contrary to the grid-based method, with the PINNs technique we choose to solve the flow equation for the derivative of the effective potential, $V_k'(\rho)$, instead of $V_k(\rho)$ itself, similarly to ~\cite{Tan2026}. The flow of $V_k'(\rho)$ is obtained from the flow equation of $V_k(\rho)$, (\ref{Eq:LPAONflow}), by simply differentiating both sides with respect to $\rho$. One may reconstruct $V_k(\rho)$ by numerically integrating $V_k'(\rho)$ with respect to $\rho$.

\begin{figure}[t]
    \centering
    \includegraphics[width=\linewidth]{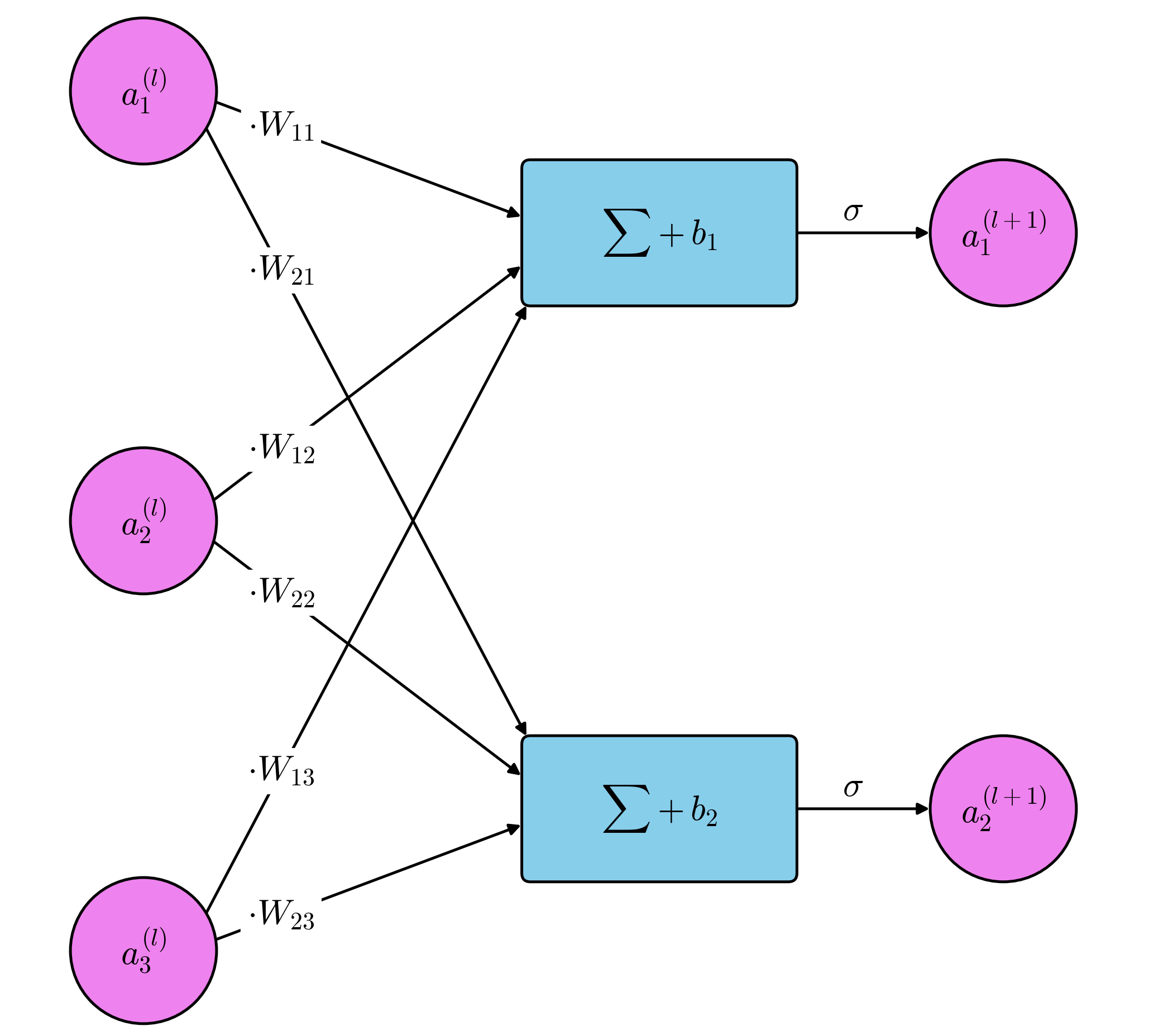}
    \caption{The nonlinear transformation connecting consecutive activation layers of a FFN.}
    \label{fig:FFN_unit_trans}
\end{figure}
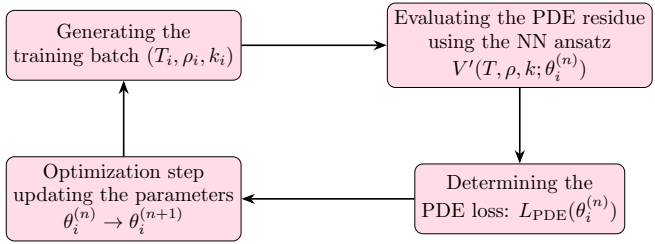
\begin{figure}[t]
	\centering
	\def\hsep{2.5cm}
	\def\vsep{1.3cm}
	\resizebox{\linewidth}{!}{%
	\begin{tikzpicture}[
		box/.style={
			draw,
			rounded corners,
			minimum width=2.5cm,
			minimum height=1.2cm,
			align=center,
			fill=cfpastelpink
		},
		arrow/.style={->, thick, >=Stealth}
		]
		
		\node[box] (data)
		{Generating the \\ training batch $(T_i, \rho_i, k_i)$};
			
		\node[box, right=\hsep of data] (evaluate)
		{Evaluating the PDE residue \\ using the NN ansatz \\ $V'(T,\rho, k; \theta_i^{(n)})$};
			
		\node[box, below=\vsep of evaluate] (loss)
		{Determining the \\ PDE loss: $L_{\text{PDE}}(\theta_i^{(n)})$};
		
		\node[box] (opt) at ($(data |- loss) + (0,0)$)
		{Optimization step \\ updating the parameters \\ $\theta_i^{(n)} \rightarrow \theta_i^{(n+1)}$};
		
		\draw[arrow] (data.east) -- (evaluate.west);
		\draw[arrow] (evaluate.south) -- (loss.north);
		\draw[arrow] (loss.west) -- (opt.east);
		\draw[arrow] (opt.north) -- (data.south);
	\end{tikzpicture}
	}
	\caption{Schematics of the PINN training. The cycle is repeated until convergence.}
\end{figure}
In this work neural networks are not used to directly approximate $V_k'(\rho)$ or calculate corrections to some truncated expansion, like the large-$N$ result. Instead, we utilize the networks as building blocks for our ansatz. For satisfying the UV boundary condition and the $\rho\to\infty$ asymptotics one may start with the following ansatz,
\bea
    \label{Eq:ansatz1}
    V_k'(\rho;T) &\!\!=\!\!& m_\Lambda^2 + \frac{g_\Lambda}{3}\rho \nonumber\\
    &\!\!+\!\!& \frac{1 - k}{\sqrt{1 + (\rho/\rho_{0,\Lambda})^3}}\mathcal{N}\left(T, \frac{\rho}{\rho_{0,\Lambda}}, k;\theta_i\right)\ ,
\eea
where $\mathcal{N}$ is a neural network in the sense of (\ref{Eq:NN}), and as before, the minimum point in the UV is $\rho_{0,\Lambda} = -3 m_\Lambda^2/g_\Lambda$. Note that with hyperbolic tangent activation, the neural network is a bounded function for fix $\theta_i$. Consequently, (\ref{Eq:ansatz1}) satisfies the $\rho\to\infty$ asymptotics. This ansatz is optimized for the ``large-$\rho$'' region, however, the phenomenon of flattening of the potential, which is the numerically most challenging part, is prominent at $\rho\lesssim \rho_{0,\Lambda}$. The (\ref{Eq:ansatz1}) ansatz, unfortunately, turns out to be insufficient to accurately capture such behavior.

The problem of the flattening of the potential can be best handled by a power series ansatz, therefore, in order to appropriately capture this region we modify (\ref{Eq:ansatz1}) as follows:
\begin{align}
    \label{Eq:ansatz2}
    V_k'(\rho;T) &= \sum_{n=0}^{P}\left(\frac{(1-k)\mathcal{N}_n(T,k; \theta_i)}{\sqrt{1 + (\rho/\rho_{0,\Lambda})^{2n+3}}} + a^{(n)}_{\text{UV}}\right)(\rho/\rho_{0,\Lambda})^n,
\end{align}
where the UV coefficients are
\begin{align}
	a_{\text{UV}}^{(n)}
	&=
	\begin{cases}
		m_\Lambda^2 & \text{if } n = 0\\
		\cfrac{g_\Lambda\rho_{0,\Lambda}}{3} &  \text{if } n = 1\\
		0 &  \text{if } n>1
	\end{cases}\ .
\end{align}
The new ansatz, (\ref{Eq:ansatz2}), behaves like an order $P$ polynomial for ``small $\rho$'', but still satisfies the $\rho\to\infty$ asymptotics as well as the UV boundary condition. Note that, the neural network is practically used to predict the coefficients in this series and thus has $P+1$ output values, $\mathcal{N}_n(T, k;\theta_i)$ ($n\in\{0, 1, \dots, P\}$). Also note that, for this ansatz the neural network does not take $\rho$ as an input, because the primary $\rho$ dependence is assumed to be handled by the already present $\rho$ dependent pieces in the ansatz. The transition between ``small'' and ``large'' $\rho$ is controlled by the minimum point of the UV potential, $\rho_{0, \Lambda}$. 

The trainings showed that while this ansatz shows significant improvements compared to the former one, it still fails to accurately capture the flattening. To tackle the problem, we further postulate a second modification:\vspace{-0.4cm}
\begin{widetext}
    \begin{align}
        \label{ansatz_semifinal}
        V_k'(\rho;T) &= \overbrace{\left(\frac{c_k^{(0)}}{\sqrt{1 + (\alpha_0(T, k;\theta_i) \rho/\rho_0)^3}} + a_{\text{UV}}^{(0)}\right)(\rho/\rho_0)^0}^{n=0} + \overbrace{\left(\frac{c_k^{(0)} a_{\text{UV}}^{(1)}/a_{\text{UV}}^{(0)} + (1 - k)\mathcal{N}_1(T, k;\theta_i)}{\sqrt{1 + (\rho/\rho_0)^2 (\alpha_1(T, k;\theta_i) \rho/\rho_0)^3}} + a_{\text{UV}}^{(1)}\right)(\rho/\rho_0)^1}^{n=1} +\nonumber\\
        &+\sum_{n=2}^{P}\left(\frac{(1-k)\mathcal{N}_n(T, k; \theta_i)}{\sqrt{1 + (\rho/\rho_0)^{2n} (\alpha_n(T, k;\theta_i) \rho/\rho_0)^3}} + a^{(n)}_{\text{UV}}\right)(\rho/\rho_0)^n\ ,
    \end{align}
\end{widetext}
where we use
\begin{align}
    c_k^{(0)} &= (1 + a_{\text{UV}}^{(0)})k^3 - k^2 - a_{\text{UV}}^{(0)} +\nonumber\\
    &+(1-k)\text{softplus}(\mathcal{N}_0(T, k;\theta_i)),
\end{align}         
with
\begin{align}
\text{softplus}(x) = \ln\left(1 + e^x\right)\ .
\end{align}
Note that now the neural network has $2(P+1)$ outputs, $\mathcal{N}_n(k;\theta_i)$ and $\alpha_n(k;\theta_i)$, where the latter provide a more flexible transition scale between the ``small'' $\rho$ and ``large'' $\rho$ regions. These scale features are not plain outputs of the neural network, but a modification of them (using softplus) that forces $\alpha_n(k;\theta_i) > 1$ to hold for all $n$ and $\theta_i$. Furthermore, the $n=0$ and $n=1$ coefficients are constructed in a different way from the higher order terms. Ignoring the $\mathcal{N}_0$ for the time being, the $n=0$ coefficient is constructed such that as $k\to0$ one has $V_k'(\rho=0) \approx -k^2$ (motivated by the IR behavior in the broken phase), while the $k^3$ term ensures that the UV potential remains intact. Of course the postulated formula is not the exact RG scale dependence, especially in the symmetric phase. This motivated the remaining correction term that contains $\mathcal{N}_0$. For the $n=1$ modifications, the motivation was not the flattening phenomenon, but to set the ratio between the $n=0$ and $n=1$ coefficients constant at $\rho=0$, which is realized when the $\mathcal{N}_1$ term is ignored. Similarly as before, the $\mathcal{N}_1$ term serves as a correction to this heuristic baseline setting. 

With the third ansatz we obtained very promising results, except for the flow of the root of $V_k'(\rho)$ [i.e., the minimum point of $V_k(\rho)$] upon comparison with the grid results. Interestingly, we found that appending $\rho/\rho_{0,\Lambda}$ to the inputs of the neural network solves this issue and thus we reached the final form of our ansatz: formally it is the same as \eqref{ansatz_semifinal}, but all $\mathcal{N}_n$ and $\alpha_n$ now depend on $\rho/\rho_{0,\Lambda}$ as well.

It is easy to see that the current form of the ansatz, with certain weights and biases, may reach a state, where the inverse propagators can become non-positive, i.e.,
\bea
k^2 + V_k'(\rho) + 2\rho V_k''(\rho) \leq 0
\eea
or 
\bea
k^2 + V_k'(\rho) \leq 0,
\eea
for some $(\rho,T,k)$ points. We call the set of these points the unstable region. If the generated training batch is such that it has samples from the unstable region, the corresponding PDE residues and hence the entire PDE loss is undefined, since non-positive quantities appear under a square root in the denominator [see (\ref{Eq:LPAONflow})], and thus the training fails. An easy solution to this problem is to mask out such samples, which allows the training to continue, however, this does not completely eliminate the issue. Approaching the unstable region from outside, one observes a diverging PDE residue on the boundary, since the denominator approaches zero there. This phenomenon makes the unmasked boundary of the unstable region (i.e., the closest unmasked points to the unstable region) dominate the PDE loss, implying that the optimizer will automatically prioritize reducing the unstable region. While naively one would applaud this result, in our experience the complete elimination of the unstable region comes at the price of the accurate description at higher RG scales.

This problem is tackled by introducing an IR cutoff $\varepsilon>0$ and genearing all $(T_i, \rho_i, k_i)$ triplets in the training batch such that $\varepsilon \leq k_i \leq 1$. This quantity determines a scale above which we aim to solve the flow equation while leaving room for remanent unstable regions at $k< \varepsilon$ RG scales. Note that maintaining accuracy at $k>\varepsilon$ while lowering $\varepsilon$ can be done by increasing the number of model parameters (i.e., width, depth, and polynom order), however, the detailed investigation of that is beyond the scope of this paper.

\section{Results}

\subsection{Zero temperature training}

First, we apply the PINNs method to the case of a fixed temperature, specifically $T=0$, since the lower the temperature, the more prominent the flattening of the potential gets. We expect that if the PINNs approach reproduces the grid results at $T=0$ we have all the reasons to anticipate the same at higher fixed temperature.

Fig. \ref{anharm_T0_eps02} compares the grid and the PINNs results from different points of view. One observes that in almost all aspects the two methods produce nearly identical curves (except for the flow of the quartic coupling, $g_k$) in the $\varepsilon \leq k \leq 1$ region. The ``instability threshold'', appearing in Fig. \ref{anharm_T0_eps02}, is defined as the largest RG scale $k_{\text{IT}}$ that the unstable region reaches. Note that this is not a hyperparameter\footnote{Hyperparameters are non-dynamical parameters, i.e., they are under our control and not that of the training process} of the training, the unstable region and thus the instability threshold depend on the model parameters, consequently, both are changing throughout the training. The $k_{\text{IT}}$ shown in Fig. \ref{anharm_T0_eps02} corresponds to the terminal model parameters. Interestingly, the final value of $k_{\text{IT}}$ happens to be very close to the $\varepsilon$ hyperparameter, that is the IR cutoff, introduced above. Recall that the optimizer forces the unstable region into the $0 \leq k < \varepsilon$ domain, which implies $k_{\text{IT}} < \varepsilon$, since if $k_{\text{IT}} \geq \varepsilon$, the optimizer forces $k_{\text{IT}}$ to decrease. However, as soon as the unstable region leaves the $\varepsilon \leq k \leq 1$ domain, it has no contribution to the loss and thus the optimizer will no longer decrease the instability threshold. In fact this is exactly what we see: the instability threshold lies slightly below the $\varepsilon = 0.2$ scale.
\newpage
\begin{widetext}
    \begin{center}
        \begin{figure}[H]
            \begin{subfigure}[t]{0.49\textwidth}
    		      \centering
    		      \includegraphics[width=\linewidth]{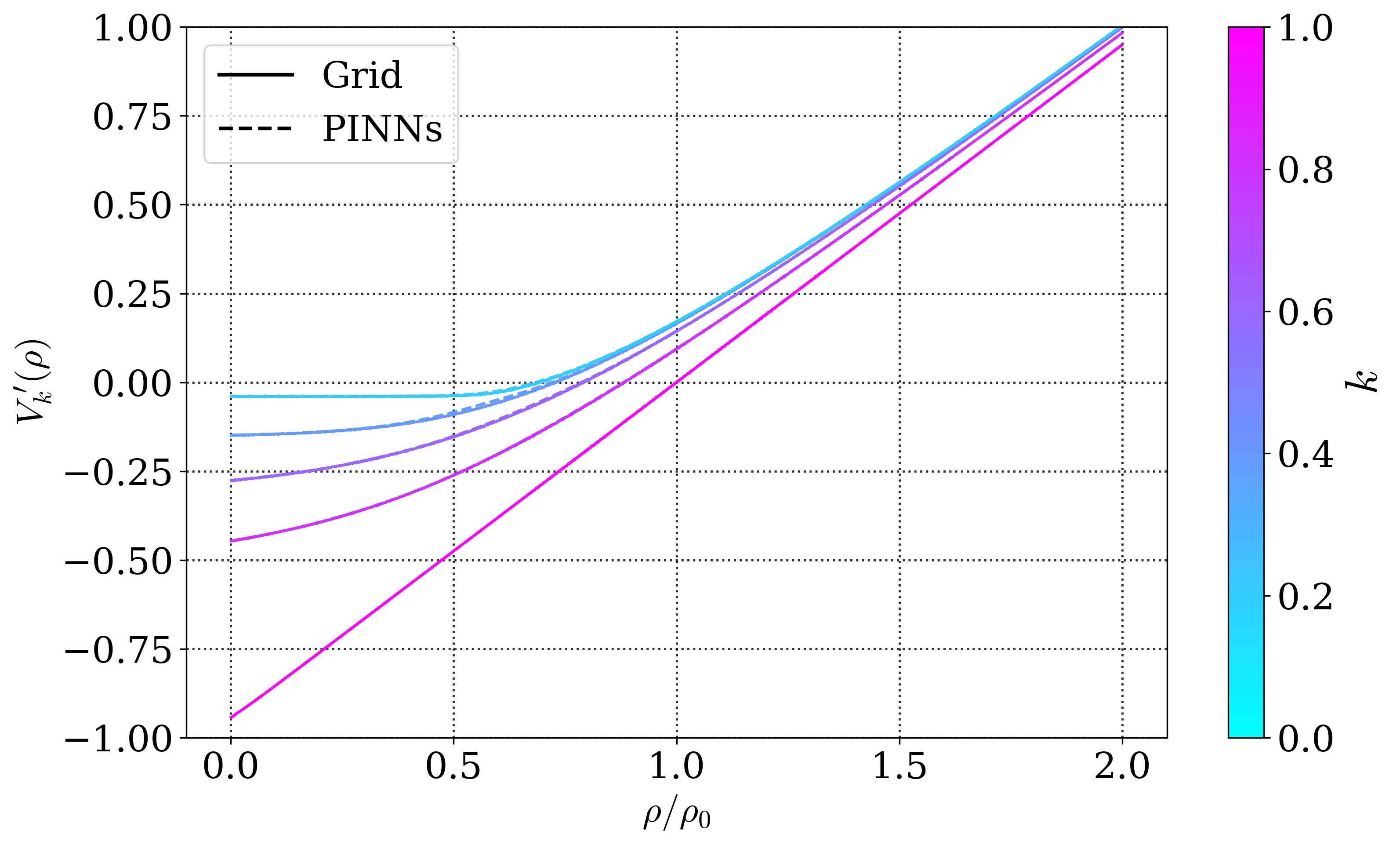}
    		      \caption{The $V_k'(\rho)$ function at different RG scales with the grid and the PINNs approaches.}
    		      \label{anharm_T0_eps02_POTCOMP}
    	    \end{subfigure}
    	      \hfill
    	    \begin{subfigure}[t]{0.49\textwidth}
    		      \centering
    		      \includegraphics[width=\linewidth]{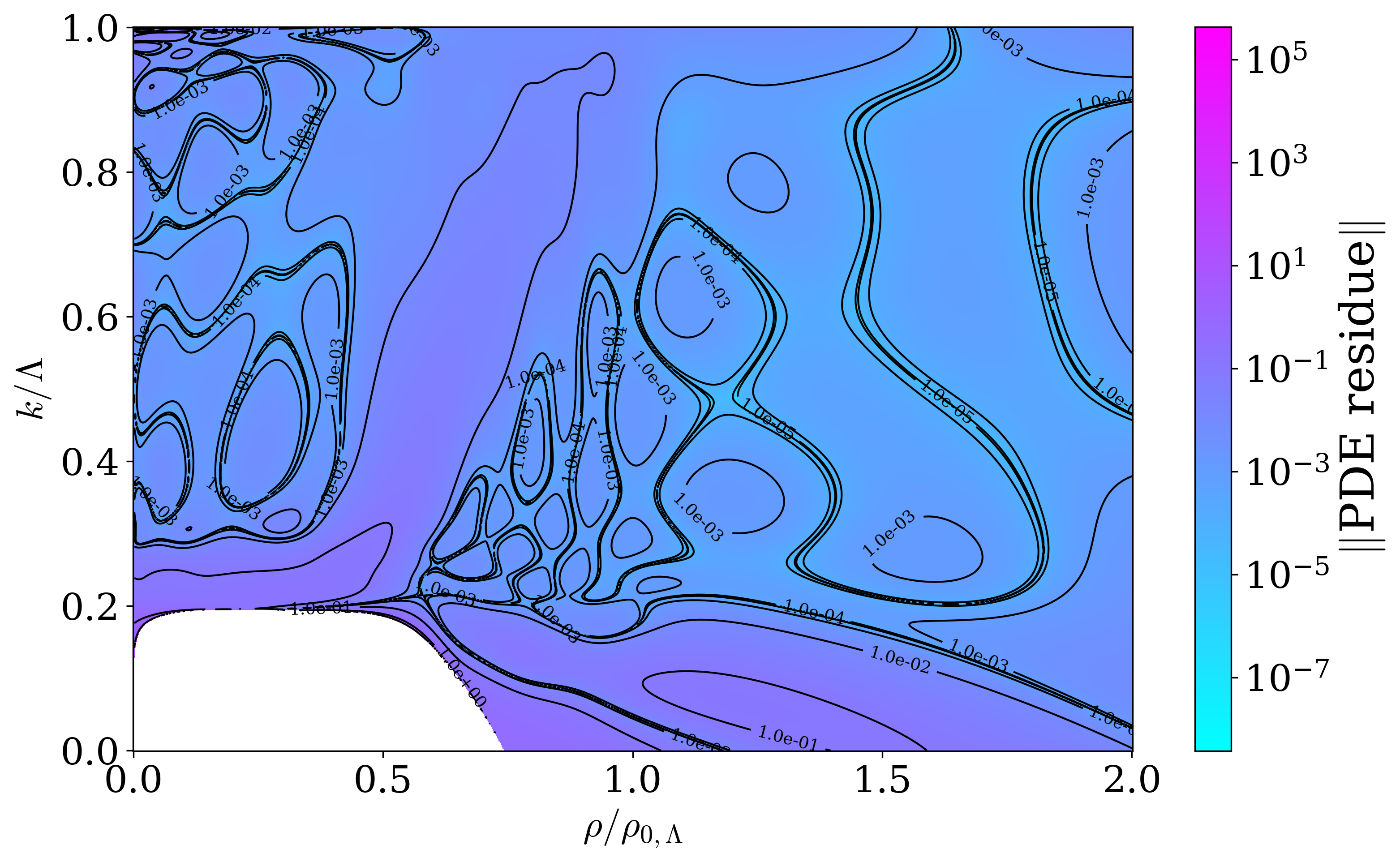}
    		      \caption{The heatmap of the PDE residue in the PINNs method in the $\rho$ -- $k$ plane. The white region is where the rhs of the flow equation is undefined, i.e., the unstable region.}
    		      \label{anharm_T0_eps02_HEATMAP}
    	    \end{subfigure}
    	    \vskip\baselineskip 
    	    \begin{subfigure}[t]{0.49\textwidth}
    		      \centering
    		      \includegraphics[width=\linewidth]{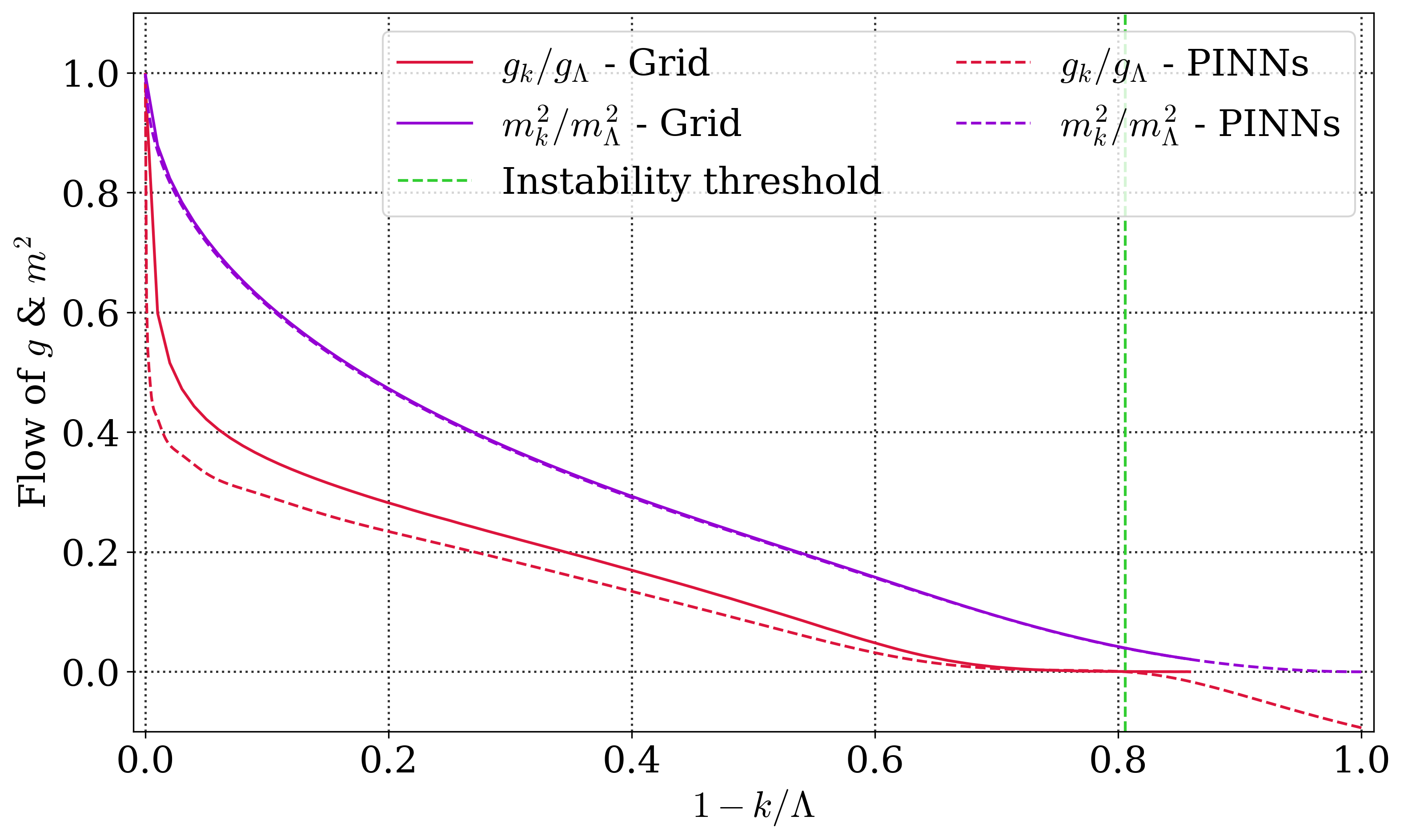}
    		      \caption{The RG flow of the mass and quartic couplings, extracted by fitting $V_k'(\rho)$ in an interval $[0,\tilde\rho]$ ($\tilde\rho\ll\rho_{0,\Lambda}$) with the line $V_k'(\rho) =m^2_k +g_k \rho/3$, both for the grid and PINNs methods.}
    		      \label{anharm_T0_eps02_M2GFLOW}
    	    \end{subfigure}
    	    \hfill
    	    \begin{subfigure}[t]{0.49\textwidth}
    		      \centering
    		      \includegraphics[width=\linewidth]{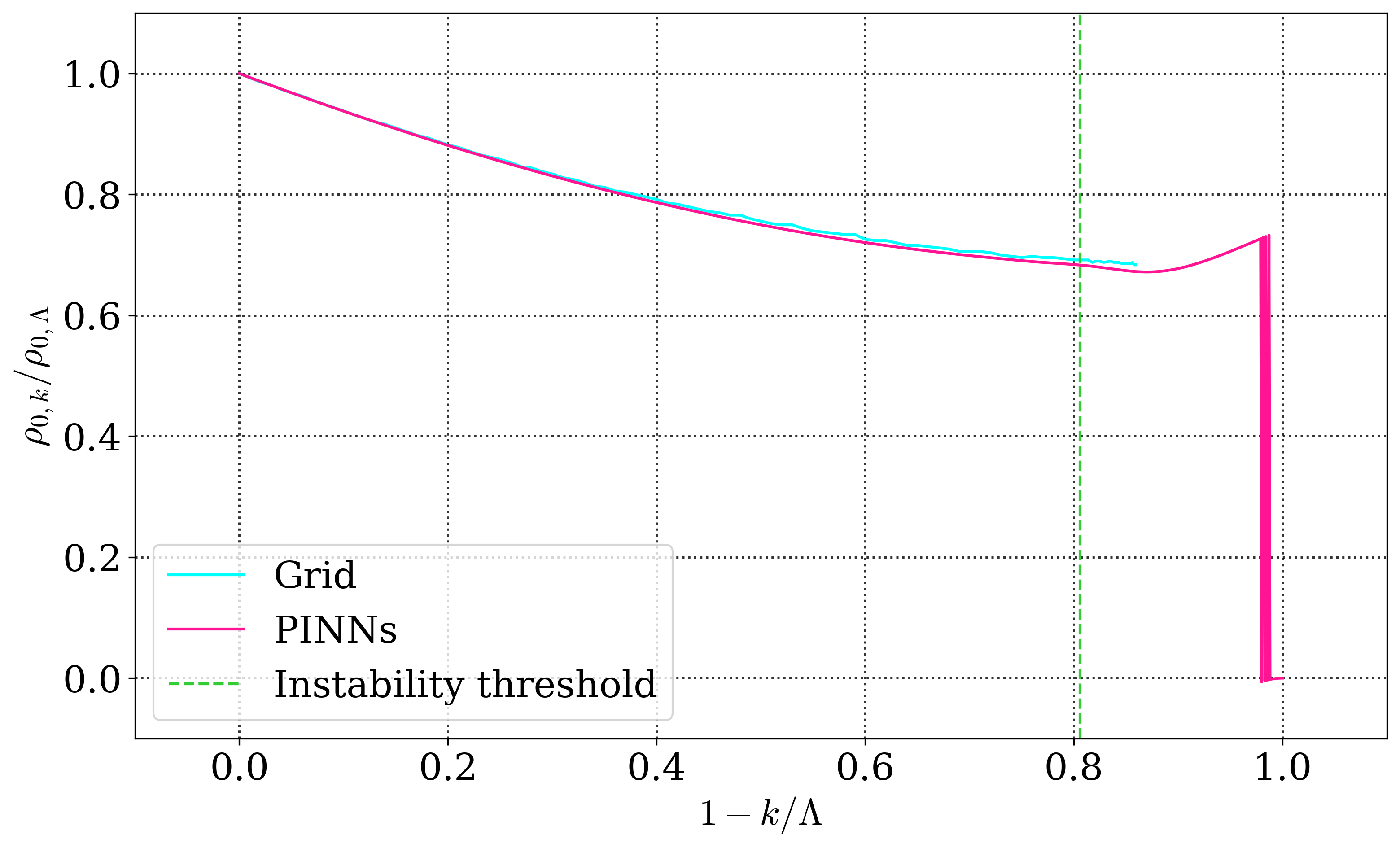}
    		      \caption{The RG flow of the root of $V_k'(\rho)$, i.e., the minimum point of $V_k(\rho)$.}
    		      \label{anharm_T0_eps02_MINFLOW}
    	    \end{subfigure}
    	    \caption{Comparing the grid and the PINNs results for the $T=0$ fixed temperature training. The results are obtained with $P=10$, $\varepsilon = 0.2$, and the neural network had 4 hidden layers with 40 neurons each.}
    	    \label{anharm_T0_eps02}
        \end{figure}
    \end{center}
\end{widetext}

\begin{figure}[h]
    \centering
    \includegraphics[width=\linewidth]{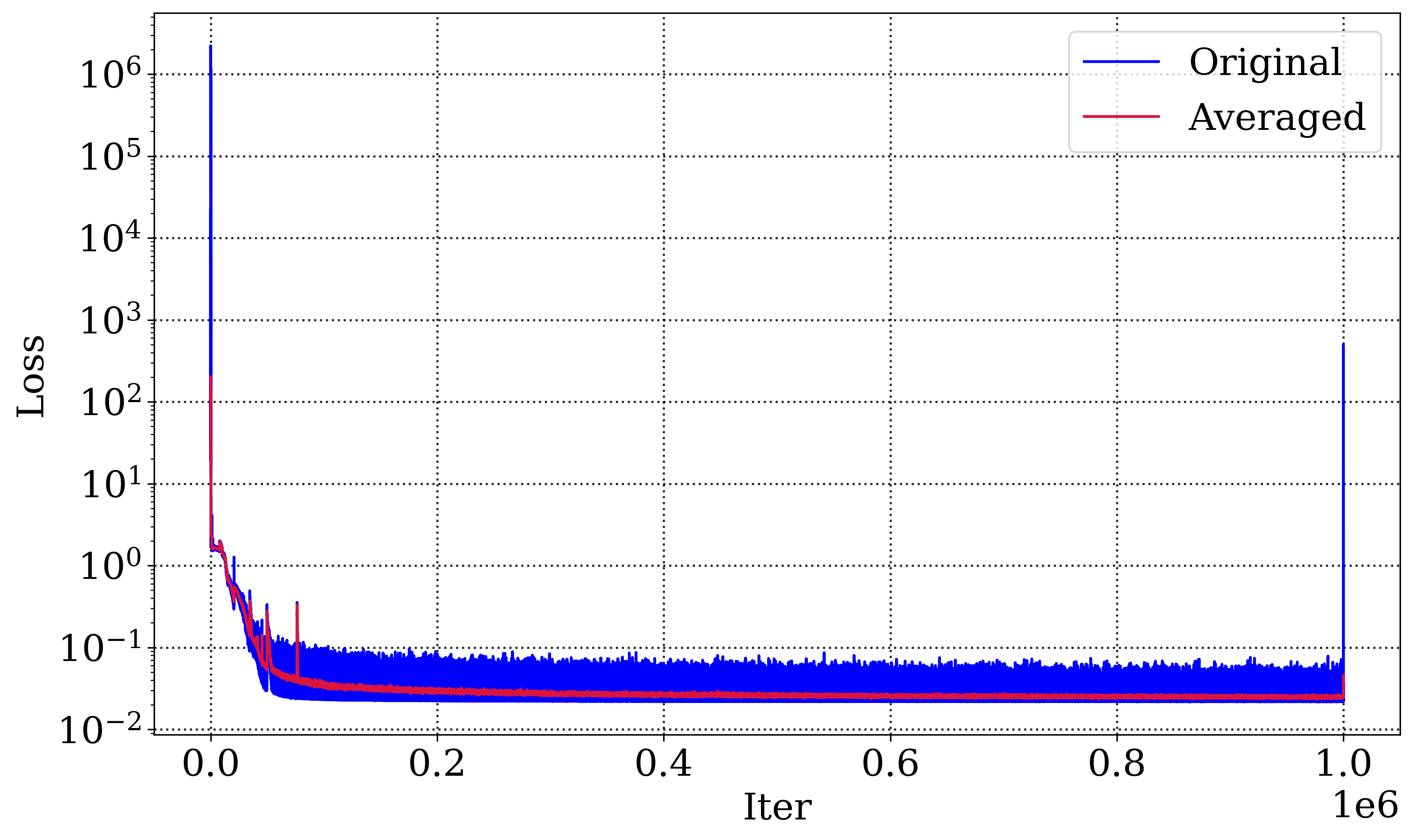}
    \caption{The loss function throughout the $T=0$ fixed temperature training.}
    \label{anharm_T0_eps02_LOSSITER}
\end{figure}

In Fig. \ref{anharm_T0_eps02_LOSSITER} we show how the loss function is being minimized throughout the training. The blue curve (\textit{Original}) shows the loss values iteration by iteration, while the red curve (\textit{Averaged}) displayes the moving logarithmic average of the former one (with a 100 iterations wide window). One can immediately see the loss being decreased, while sharp peaks are emerging on top of this baseline. This is especially true in the beginning and in the end of the flow.
As we already pointed out, the unmasked boundary of the unstable region gives a major contribution to the loss and thus when the unstable region enters the $\varepsilon\leq k\leq 1$ domain, it creates a spike in the loss--iter function. Note that the largest spike is in the beginning, which is understandable, since the weights and biases of the neural network are randomly initialized and thus it is likely going to produce a massive unstable region that spans to the $\varepsilon\leq k\leq 1$ domain. Consequently, the initial iterations are spent on extinguishing the unstable region from this domain, which results in the quick initial drop of the loss function. Up until iteration $\sim \!10^5$ we still see a couple of spikes, although far smaller than the initial one. This phenomenon can be explained by the unstable region wiggling under the $k=\varepsilon$ line and sometimes it reaches the $\varepsilon \leq k \leq 1$ domain, creating a sharp rise in the loss. These are immediately suppressed by the optimizer resulting in a sharp fall of the loss. The two sharp jumps together create what we interpret as a spike. Crucially, this phenomenon stops after iteration $\sim \!10^5$, which is a result of the decrease in the learning rate of the optimizer. In doing so, we scale down the wiggling of the unstable region and thus confining it in the $0\leq k< \varepsilon$ domain. This still does not explain the last peak right at the end, which is the result of the second order L-BFGS optimization step. It performs line searches by varying the model parameters significantly. Consequently, in the beginning of the L-BFGS optimization, the unstable region reenters the $\varepsilon \leq k \leq 1$ domain once more and thus create a final spike.

\subsection{Finite temperature training}

The next step in applying the PINNs approach to the LPA flow is to start varying the temperature. In the context of neural networks, the temperature dependence can be explored in two different ways:
\begin{itemize}
    \item For each temperature value of our interest, we train a model with fixed temperature, as described in the previous subsection.
    \item We train one model that learns the temperature dependence as well, i.e., we let the temperatures in the training batch vary.
\end{itemize}

\begin{figure}[t]
    \centering
    \includegraphics[width=\linewidth]{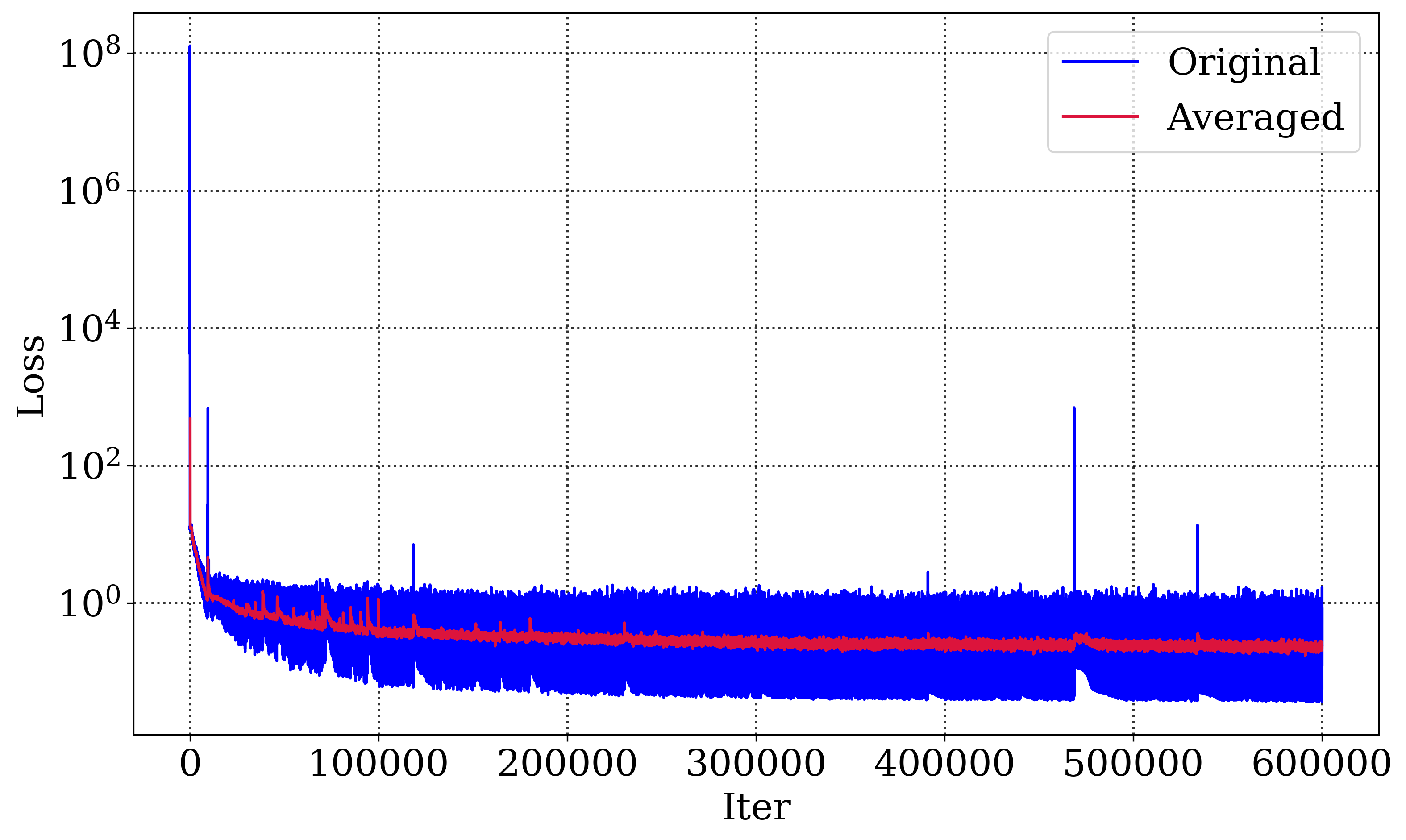}
    \caption{The loss function throughout the varying temperature training.}
    \label{anharm_varyT_HS40_eps02_LOSSITER}
\end{figure}

Compared to Ref. \cite{Tan2026}, we choose the more general second approach. We use the same neural network architecture that we used for the $T=0$ fixed temperature training, i.e., we start the exploration with the network of $\text{D} = 4$ (\textit{depth}) and $\text{HS}=40$ (\textit{hidden size}).

Fig. \ref{anharm_T0_eps02_LOSSITER} shows the loss function with increasing iteration number for the previous, $T=0$ training, while Fig. \ref{anharm_varyT_HS40_eps02_LOSSITER} displays the same for the training with varying temperature. Upon comparing the two figures, one can immediately conclude that the same model, i.e., a neural network with the same depth, hidden size, $P$, $\varepsilon$ reaches more than one order of magnitude higher terminal loss when used in the varying temperature training instead of the fixed temperature one. Of course this does not come as a surprise, it aligns well with our naive expectations coming from the fact that an additional variable is added to the training.

To further analyze the training with varying temperature, we may check its $T=0$ predictions and compare them to the results of the fixed temperature training. This can be done by analyzing Fig. \ref{anharm_varyT_HS40_eps02}. Comparing Fig. \ref{anharm_varyT_HS40_eps02_POTCOMP}, Fig. \ref{anharm_varyT_HS40_eps02_M2GFLOW} and Fig. \ref{anharm_varyT_HS40_eps02_MINFLOW} to their corresponding fixed temperature plots one draws the conclusion that indeed, the $T=0$ precision is partially sacrificed in order to describe the temperature dependence. 

Fig. \ref{anharm_varyT_HS40_eps02_HEATMAP} shows that the shape of the unstable region is modified and its area shrinks. One also notes that in the $\varepsilon\leq k \leq 1$ domain the PDE residue heatmap is significantly less fragmented compared to Fig. \ref{anharm_T0_eps02_HEATMAP}. In princple, the fragmentation is the consequence of the PDE residue rapidly oscillating around zero, therefore, the lack of fragmentation in Fig. \ref{anharm_varyT_HS40_eps02_HEATMAP} is another sign of poorer overall results.

Fig. \ref{anharm_varyT_HS40_eps02_MINPOS_K_T_DEP} displays the running of the minimum point of the effective potential for different temperatures, while Fig. \ref{anharm_varyT_HS40_eps02_MINPOS_T_DEP} shows the corresponding extrapolations towards the deep IR, i.e., $k\rightarrow 0$. Here we note that the extrapolation used to determine the minimum point, $\rho_{0, k=0}$, was carried out by fitting the $k\mapsto\rho_{0, k}$ map at a given temperature by
\begin{align}
    \label{RG_flow_extrap}
	\rho_0^{(\text{fit})}(k; A, p, C) = \text{ReLU}\left(C + A k^p\right)\ ,
\end{align}
where $\text{ReLU}(x)=\text{max}\{0, x\}$. Therefore, by obtaining the best fits for $A^*$, $p^*$ and $C^*$, the result of the extrapolation is $\rho_{0, k=0} = \text{ReLU}(C^*)$. Note that, if $C^*<0$, then $\rho_{0, k=0}=0$ automatically. Repetition of this process for different temperatures yields to Fig. \ref{anharm_varyT_HS40_eps02_MINPOS_T_DEP}.

All in all, by comparing the PINNs results at this point to that of the grid method, we may conclude that the $\text{HS} = 40$ neural network is not enough to accurately represent the temperature dependence on top of the $\rho$ and $k$ dependencies. However, it is clear that Fig. \ref{anharm_varyT_HS40_eps02_MINPOS_K_T_DEP} and Fig. \ref{anharm_varyT_HS40_eps02_MINPOS_T_DEP} show promising trends. 

\begin{widetext}
    \begin{center}
        \begin{figure}[H]
            \begin{subfigure}[t]{0.49\textwidth}
    		      \centering
    		      \includegraphics[width=\linewidth]{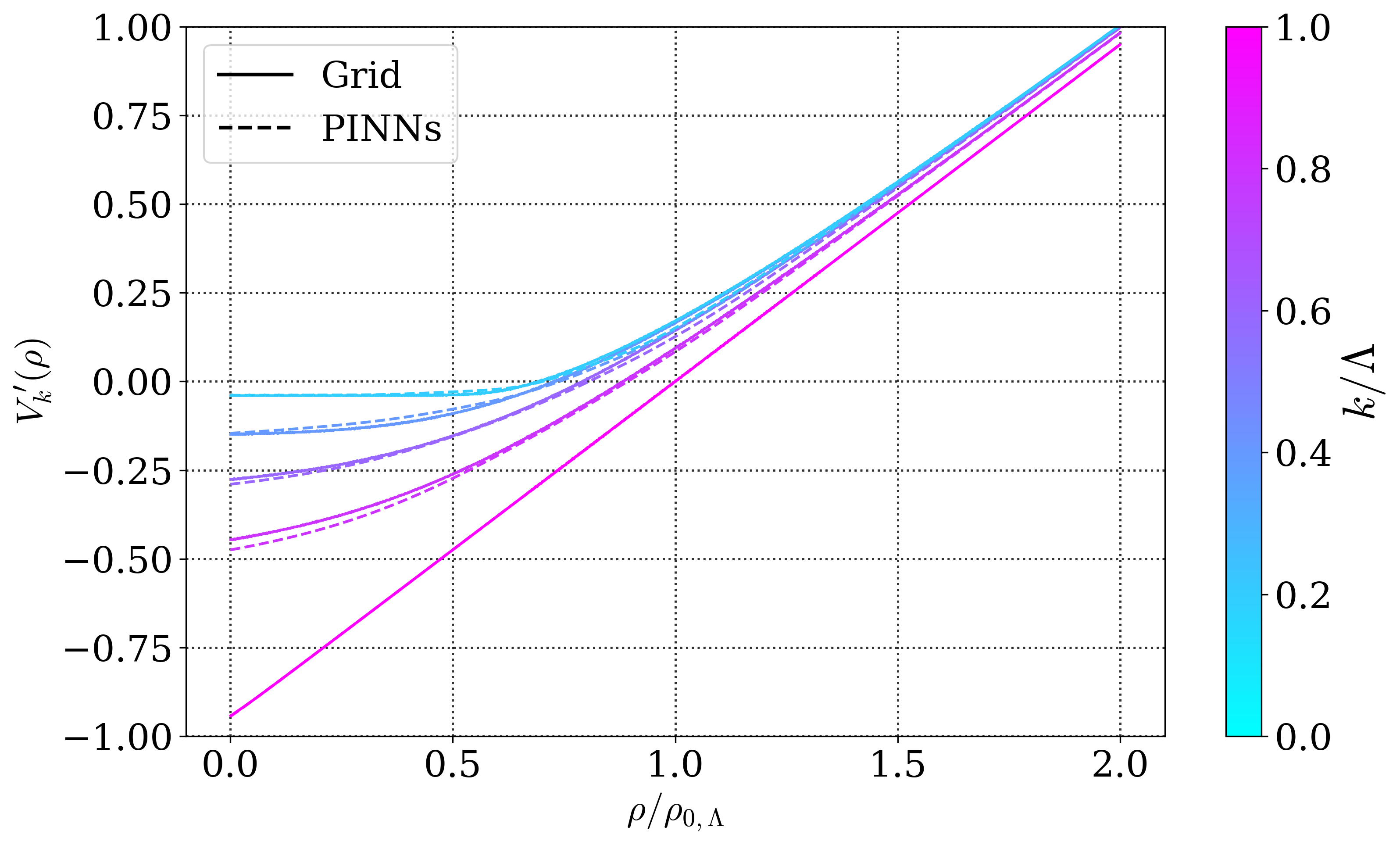}
    		      \caption{The zero temperature $V_k'(\rho)$ function at different RG scales with the grid method and the PINNs approach with varying temperature training.}
    		      \label{anharm_varyT_HS40_eps02_POTCOMP}
    	    \end{subfigure}
    	      \hfill
    	    \begin{subfigure}[t]{0.49\textwidth}
    		      \centering
    		      \includegraphics[width=\linewidth]{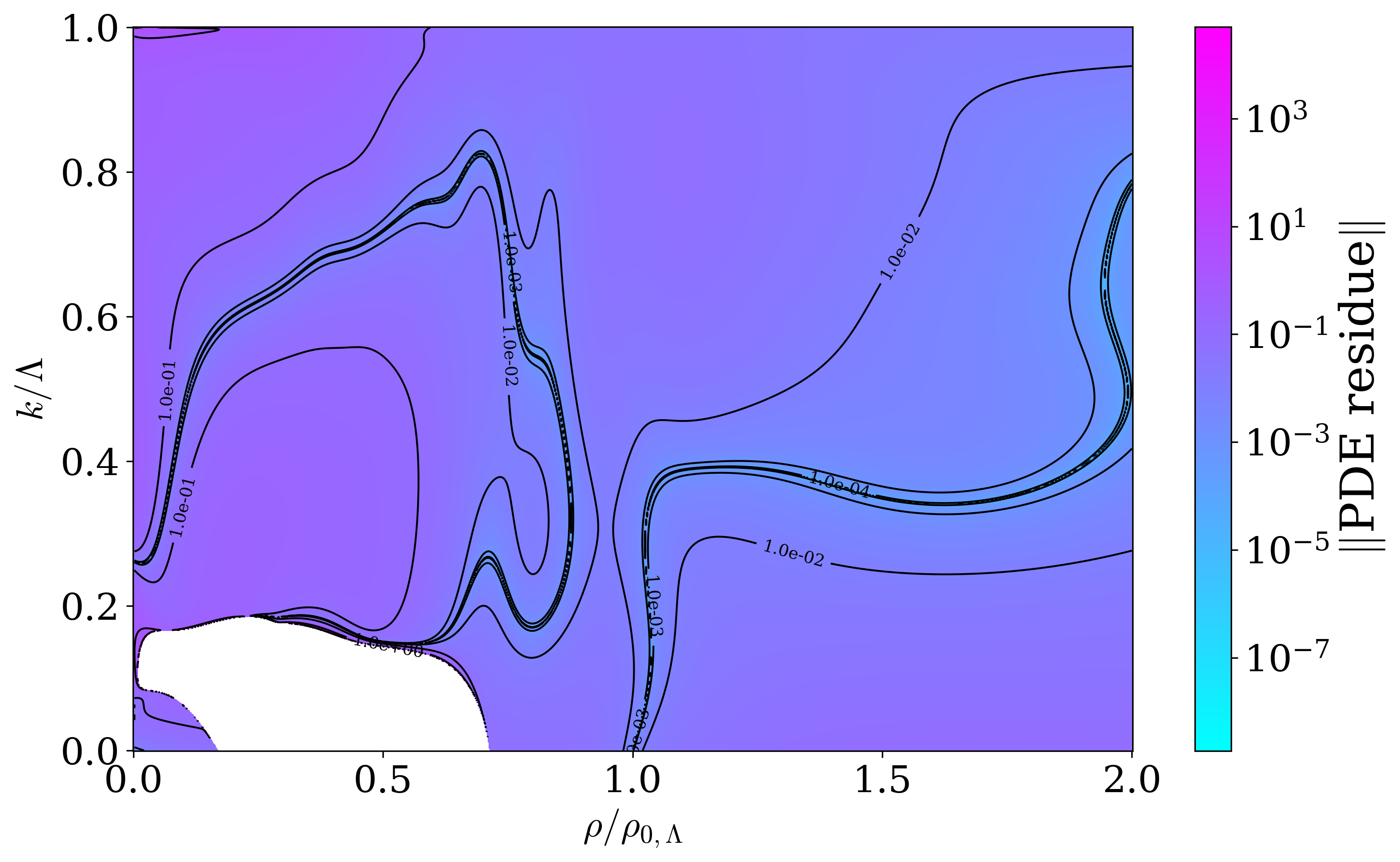}
    		      \caption{The heatmap of the PDE residue in the PINNs method with varying temperature training in the $\rho$ -- $k$ plane for zero temperature. The white region is where the rhs of the flow equation is undefined, i.e., the unstable region.}
    		      \label{anharm_varyT_HS40_eps02_HEATMAP}
    	    \end{subfigure}
    	    \vskip\baselineskip 
    	    \begin{subfigure}[t]{0.49\textwidth}
    		      \centering
    		      \includegraphics[width=\linewidth]{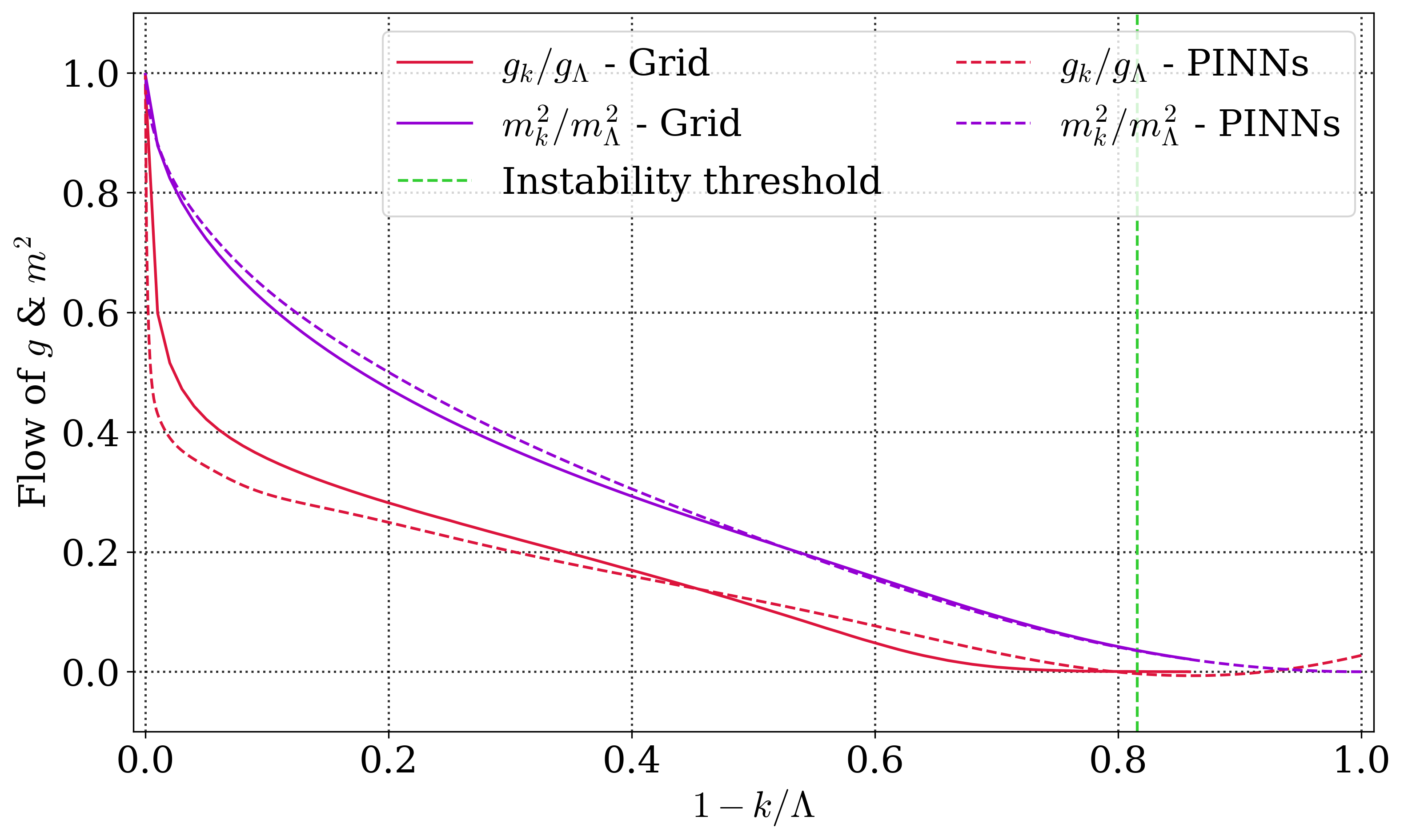}
    		      \caption{The RG flow of Taylor coefficients at $T=0$, extracted by fitting $V_k'(\rho)$ in an interval $[0,\tilde\rho]$ ($\tilde\rho\ll\rho_0$) with the line $V_k'(\rho) =m^2_k +g_k \rho/3$, both for the grid and PINNs methods.}
    		      \label{anharm_varyT_HS40_eps02_M2GFLOW}
    	    \end{subfigure}
    	    \hfill
    	    \begin{subfigure}[t]{0.49\textwidth}
    		      \centering
    		      \includegraphics[width=\linewidth]{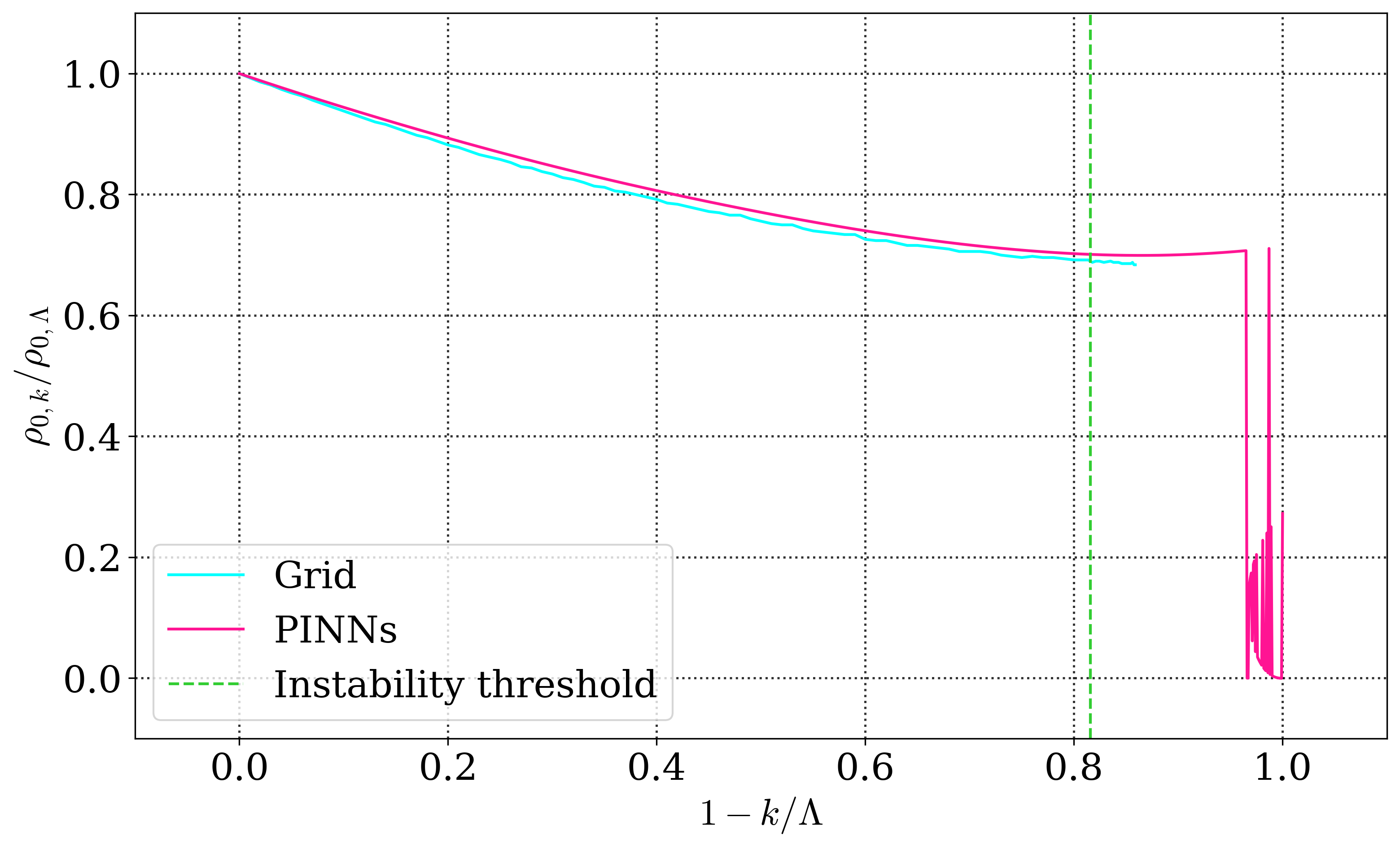}
    		      \caption{The RG flow of the root of $V_k'(\rho)$, i.e., the minimum point of $V_k(\rho)$ at $T=0$.}
    		      \label{anharm_varyT_HS40_eps02_MINFLOW}
    	    \end{subfigure}
            \vskip\baselineskip 
            \begin{subfigure}[t]{0.49\textwidth}
    		      \centering
    		      \includegraphics[width=\linewidth]{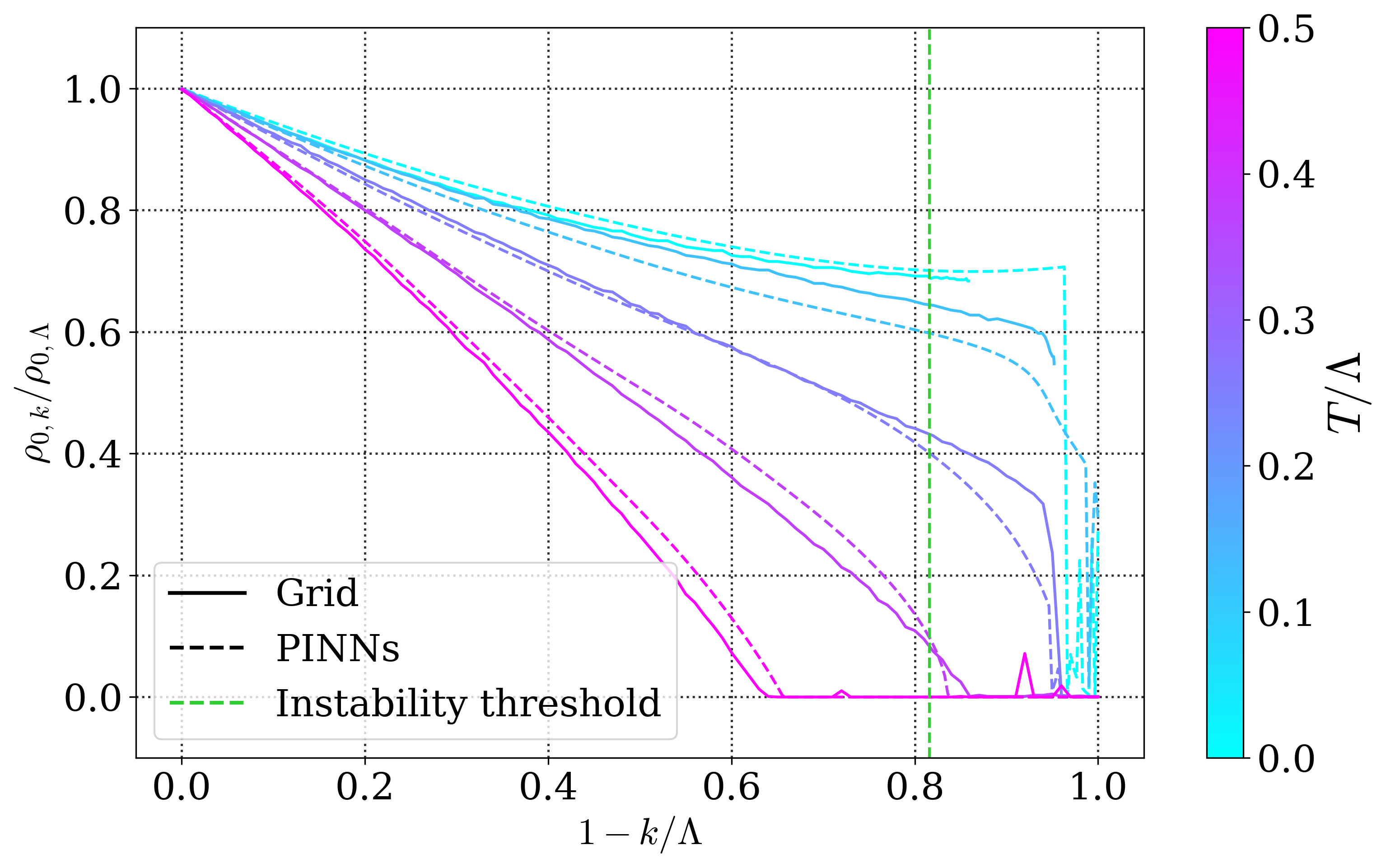}
    		      \caption{The RG flow of the root of $V_k'(\rho)$, i.e., the minimum point of $V_k(\rho)$ at different temperatures.}
    		      \label{anharm_varyT_HS40_eps02_MINPOS_K_T_DEP}
    	    \end{subfigure}
    	      \hfill
    	    \begin{subfigure}[t]{0.49\textwidth}
    		      \centering
    		      \includegraphics[width=\linewidth]{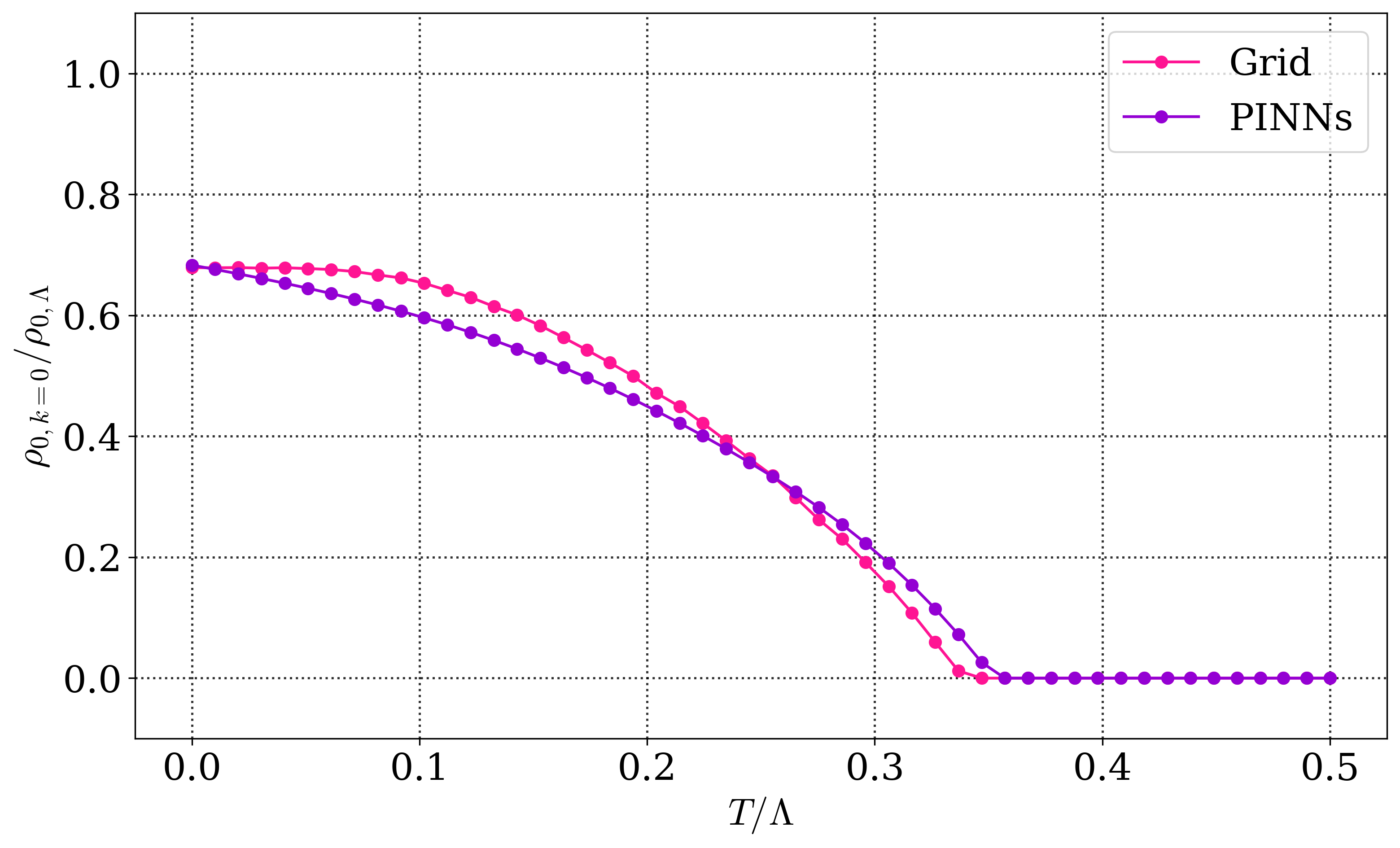}
    		      \caption{The extrapolated $\rho_{0,k=0}$ as a function of the temperature. Note the phase transition occurs at $T\sim 0.33\Lambda$.}
    		      \label{anharm_varyT_HS40_eps02_MINPOS_T_DEP}
    	    \end{subfigure}
    	    \caption{Comparing the grid and the PINNs results with varying temperature training. The parameters of the network are $P=10$, $\varepsilon = 0.2$, $D=4$ and $\text{HS} = 40$. The purpose of these plots is to compare the fixed temperature training and the varying temperature training on equal footing and not to showcase the final results that we achieved with the varying temperature training.}
    	    \label{anharm_varyT_HS40_eps02}
        \end{figure}
    \end{center}
\end{widetext}

\subsection{Improved finite temperature training}

It is reasonable to assume that a training with increased hidden size would yield better results, therefore, in this subsection we start to experiment with larger $\text{HS}$ models. Indeed, Fig. \ref{anharm_varyT_varyHS} shows that the prediction of the PINNs method for the RG flow of the minimum at different temperatures gets closer to the grid results as we go to higher in $\text{HS}$. Interestingly, this is not only true at the $\varepsilon \leq k \leq 1$ domain ($\varepsilon=0.2)$, which is randomly sampled and thus taught during training, but also at scales below $\varepsilon$. It appears that the model generalizes its knowledge from the $\varepsilon \leq k \leq 1$ domain to parts of the $0 \leq k \leq \varepsilon$ region as well. This signals that the choice of the ansatz indeed aligns well with the problem at hand.

Our point on the more general knowledge of the network is further strengthened by Fig. \ref{anharm_varyT_varyHS_PDE_RES_HM}. It shows that the unstable region, confined in the $0\leq k\leq \varepsilon$ domain, is shrinking as we increase $\text{HS}$. This is highly unexpected, since the unstable region is confined outside of the training domain and hence it is practically ``invisible'' for the optimizer. Consequently, the shrinking of the unstable region is yet another proof that the model can partially generalize beyond the training domain, moreover, it does it better as $\text{HS}$ is increased.

While it is reassuring to see the tendencies of generalized knowledge, the reason we started to increase $\text{HS}$ is to increase the number of degrees of freedom of the ansatz, hoping that the model has the capacity to accurately represent the $T$-dependence as well. Fig. \ref{anharm_varyT_varyHS} already showed that results become more accurate with increasing HS, but the learned $T$-dependence can also be illustrated via the extrapolated $\rho_{0,k=0}(T)$ curves [see (\ref{RG_flow_extrap})] at different hidden sizes, shown in Fig. \ref{anharm_varyT_MINPOS_ANAL}.

\begin{widetext}
    \begin{center}
        \begin{figure}[H]
            \begin{subfigure}[t]{0.49\textwidth}
    		      \centering
    		      \includegraphics[width=\linewidth]{varying_T_HS40_all/MinPos_k_and_T_dep.png}
    		      \caption{$\text{HS} = 40$}
    		      \label{anharm_varyT_eps02_MIN_HS40}
    	    \end{subfigure}
    	      \hfill
    	    \begin{subfigure}[t]{0.49\textwidth}
    		      \centering
    		      \includegraphics[width=\linewidth]{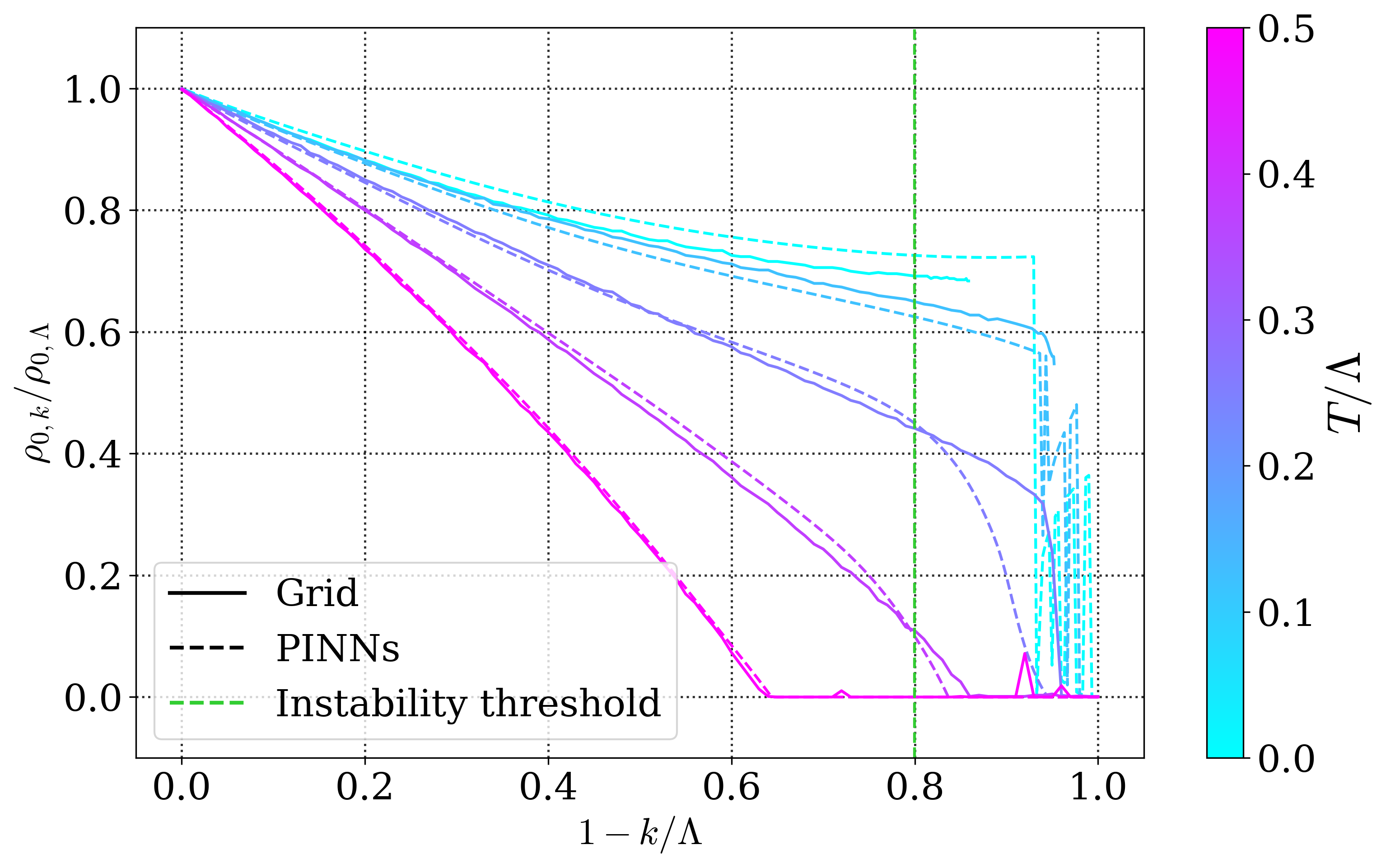}
    		      \caption{$\text{HS} = 50$}
    		      \label{anharm_varyT_eps02_MIN_HS50}
    	    \end{subfigure}
    	    \vskip\baselineskip 
    	    \begin{subfigure}[t]{0.49\textwidth}
    		      \centering
    		      \includegraphics[width=\linewidth]{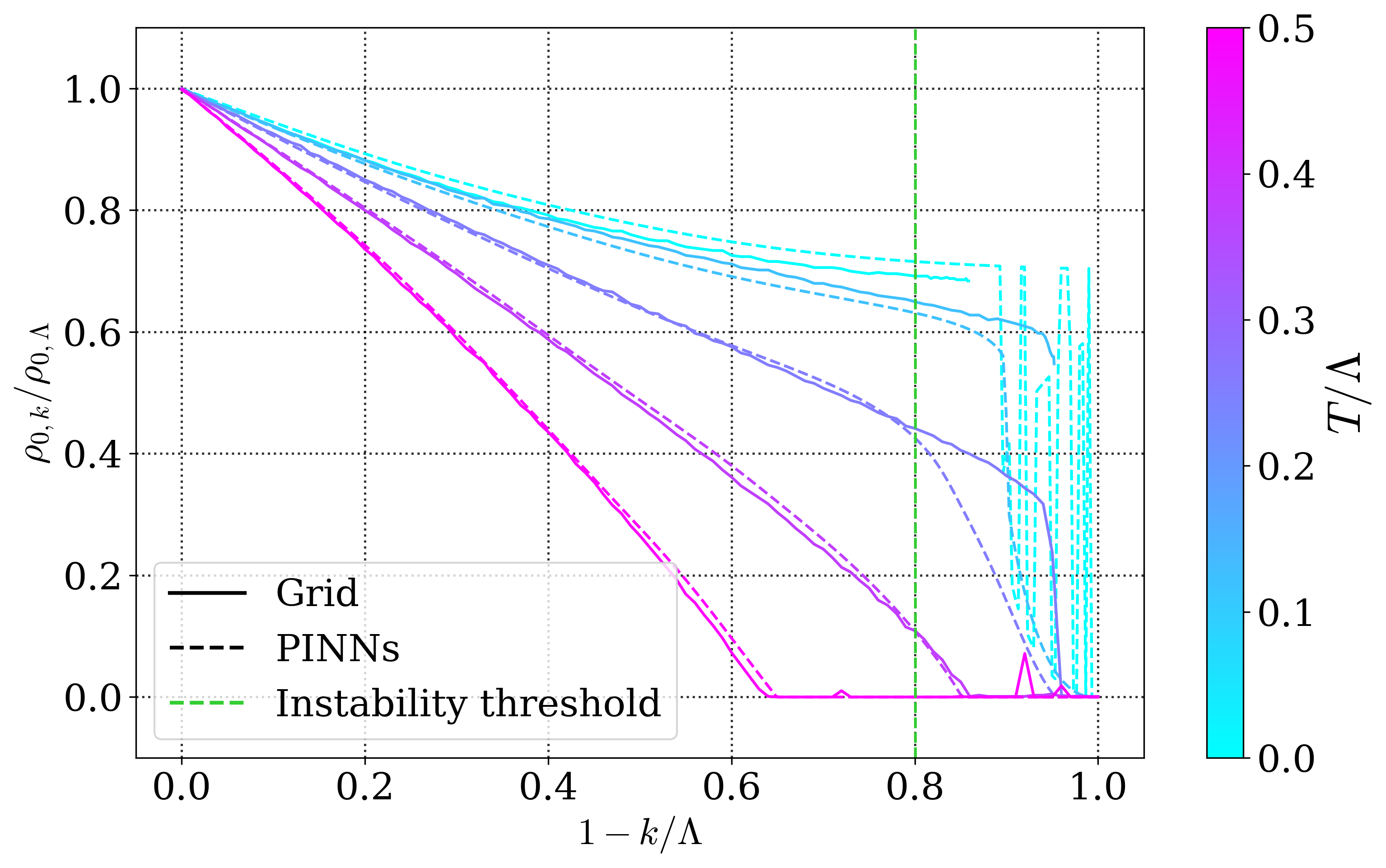}
    		      \caption{$\text{HS} = 60$}
    		      \label{anharm_varyT_eps02_MIN_HS60}
    	    \end{subfigure}
    	    \hfill
    	    \begin{subfigure}[t]{0.49\textwidth}
    		      \centering
    		      \includegraphics[width=\linewidth]{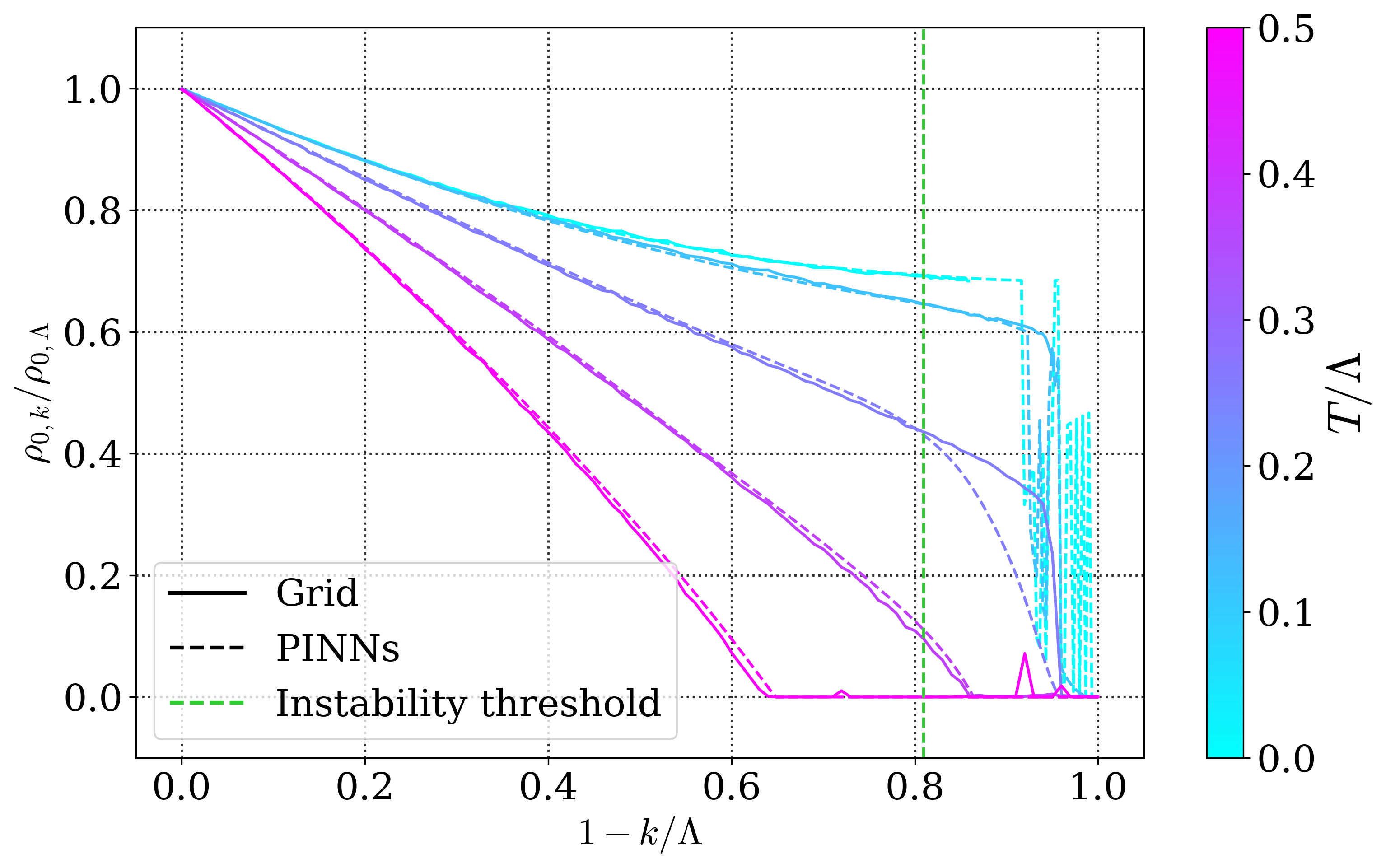}
    		      \caption{$\text{HS} = 100$}
    		      \label{anharm_varyT_eps02_MIN_HS100}
    	    \end{subfigure}
    	    \caption{The RG flow of the root of $V_k'(\rho)$ [i.e., the minimum point of $V_k(\rho)$] at different temperatures for increasing hidden sizes.}
    	    \label{anharm_varyT_varyHS}
        \end{figure}
    \end{center}

    \begin{center}
        \begin{figure}[H]
            \begin{subfigure}[t]{0.49\textwidth}
    		      \centering
    		      \includegraphics[width=\linewidth]{varying_T_HS40_all/PDE_residue_heatmap.png}
    		      \caption{$\text{HS} = 40$}
    		      \label{anharm_varyT_eps02_MIN_HS40}
    	    \end{subfigure}
    	      \hfill
    	    \begin{subfigure}[t]{0.49\textwidth}
    		      \centering
    		      \includegraphics[width=\linewidth]{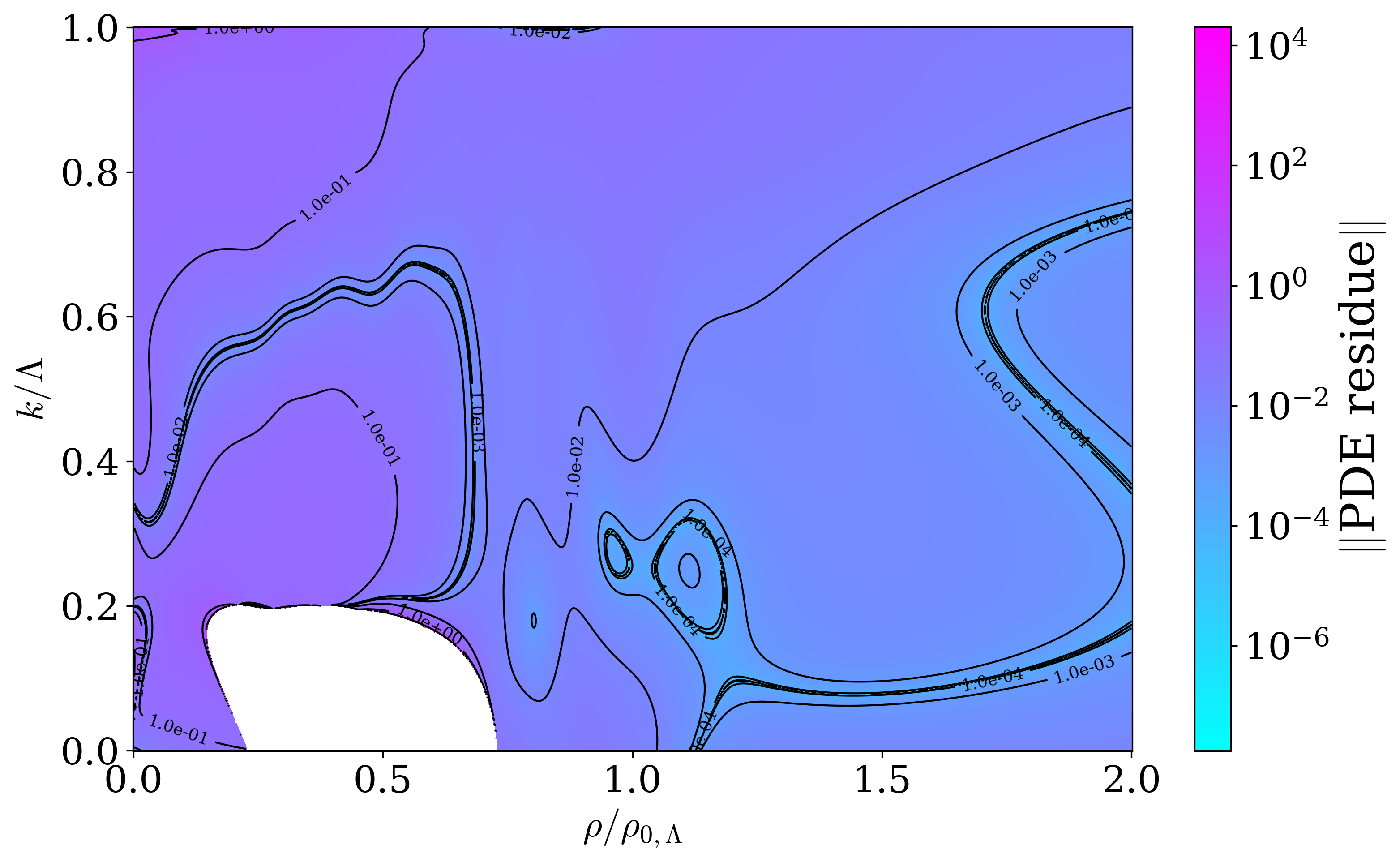}
    		      \caption{$\text{HS} = 50$}
    		      \label{anharm_varyT_eps02_MIN_HS50}
    	    \end{subfigure}
    	    \vskip\baselineskip 
    	    \begin{subfigure}[t]{0.49\textwidth}
    		      \centering
    		      \includegraphics[width=\linewidth]{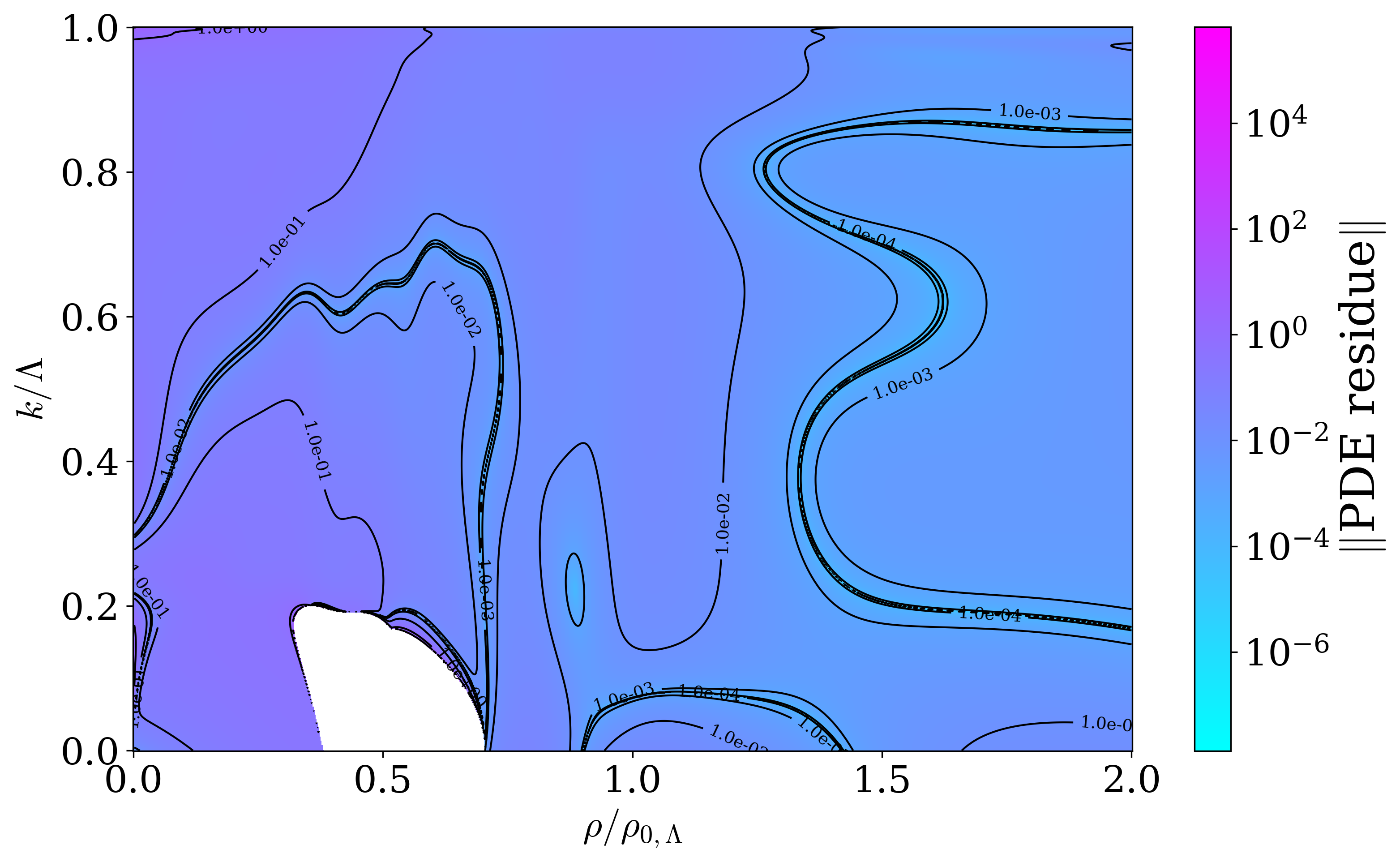}
    		      \caption{$\text{HS} = 60$}
    		      \label{anharm_varyT_eps02_MIN_HS60}
    	    \end{subfigure}
    	    \hfill
    	    \begin{subfigure}[t]{0.49\textwidth}
    		      \centering
    		      \includegraphics[width=\linewidth]{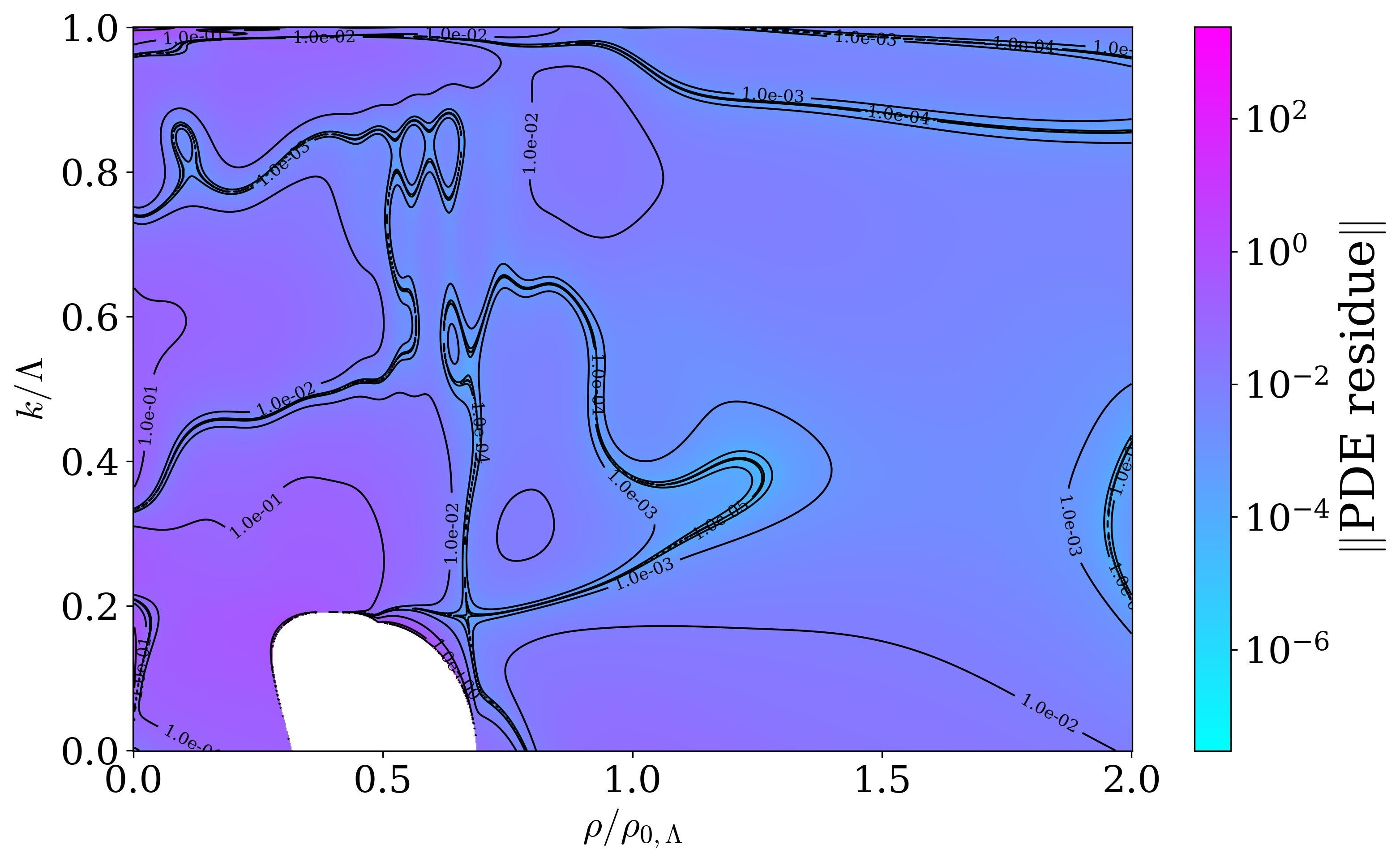}
    		      \caption{$\text{HS} = 100$}
    		      \label{anharm_varyT_eps02_MIN_HS100}
    	    \end{subfigure}
    	    \caption{The terminal PDE residue heatmaps from the varying temperature trainings, evaluated at $T=0$ for increasing hidden sizes.}
    	    \label{anharm_varyT_varyHS_PDE_RES_HM}
        \end{figure}
    \end{center}

    \begin{center}
        \begin{figure}[H]
            \begin{subfigure}[t]{0.49\textwidth}
    		      \centering
    		      \includegraphics[width=\linewidth]{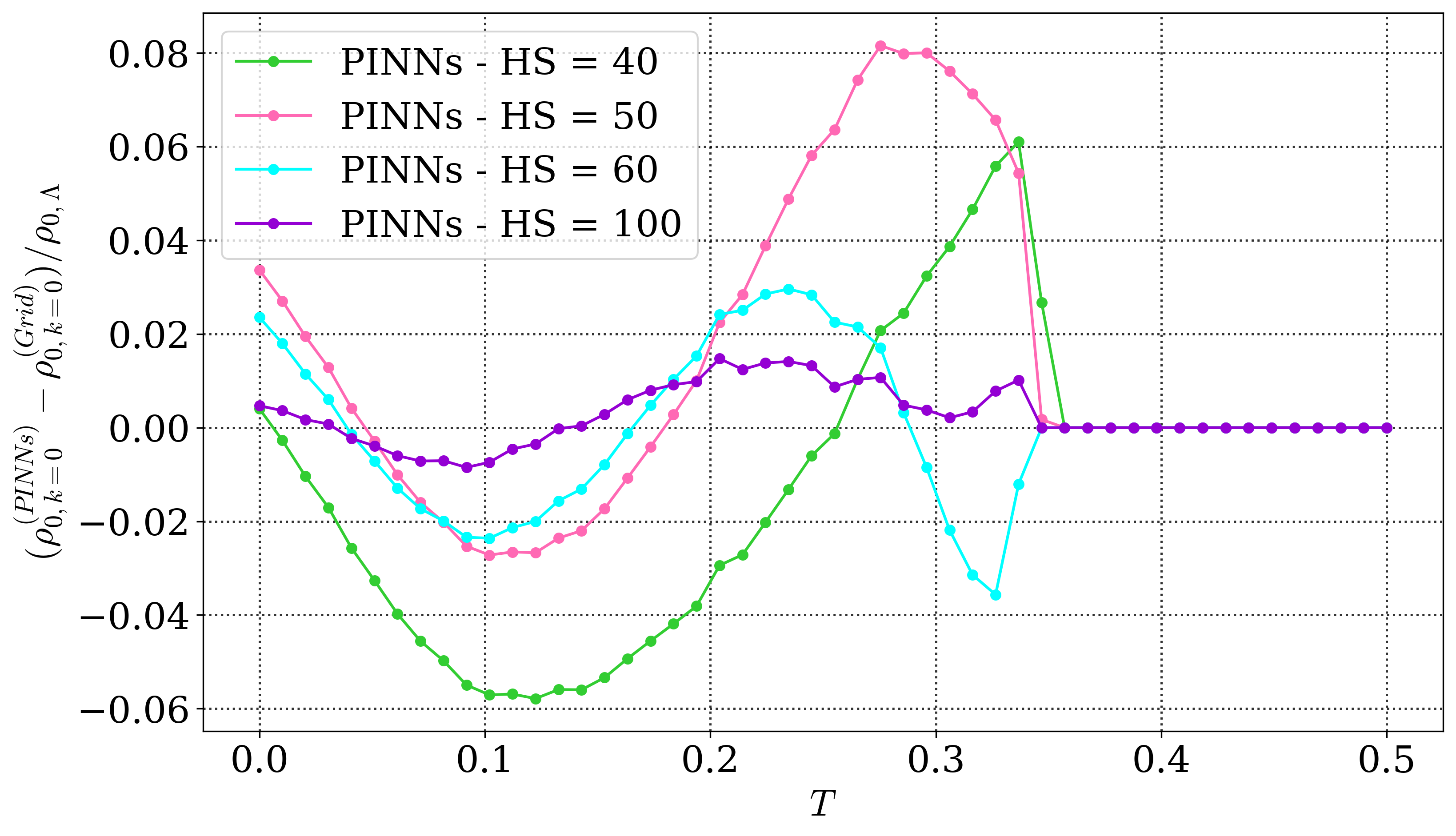}
    		      \caption{Difference between the extrapolated $\rho_{0,k=0}(T)$ results via the grid and PINNs, as a function of $T$ for different $\text{HS}$ values.}
    		      \label{anharm_varyT_varyHS_MINPOS_DIFF}
    	    \end{subfigure}
    	      \hfill
    	    \begin{subfigure}[t]{0.49\textwidth}
    		      \centering
    		      \includegraphics[width=\linewidth]{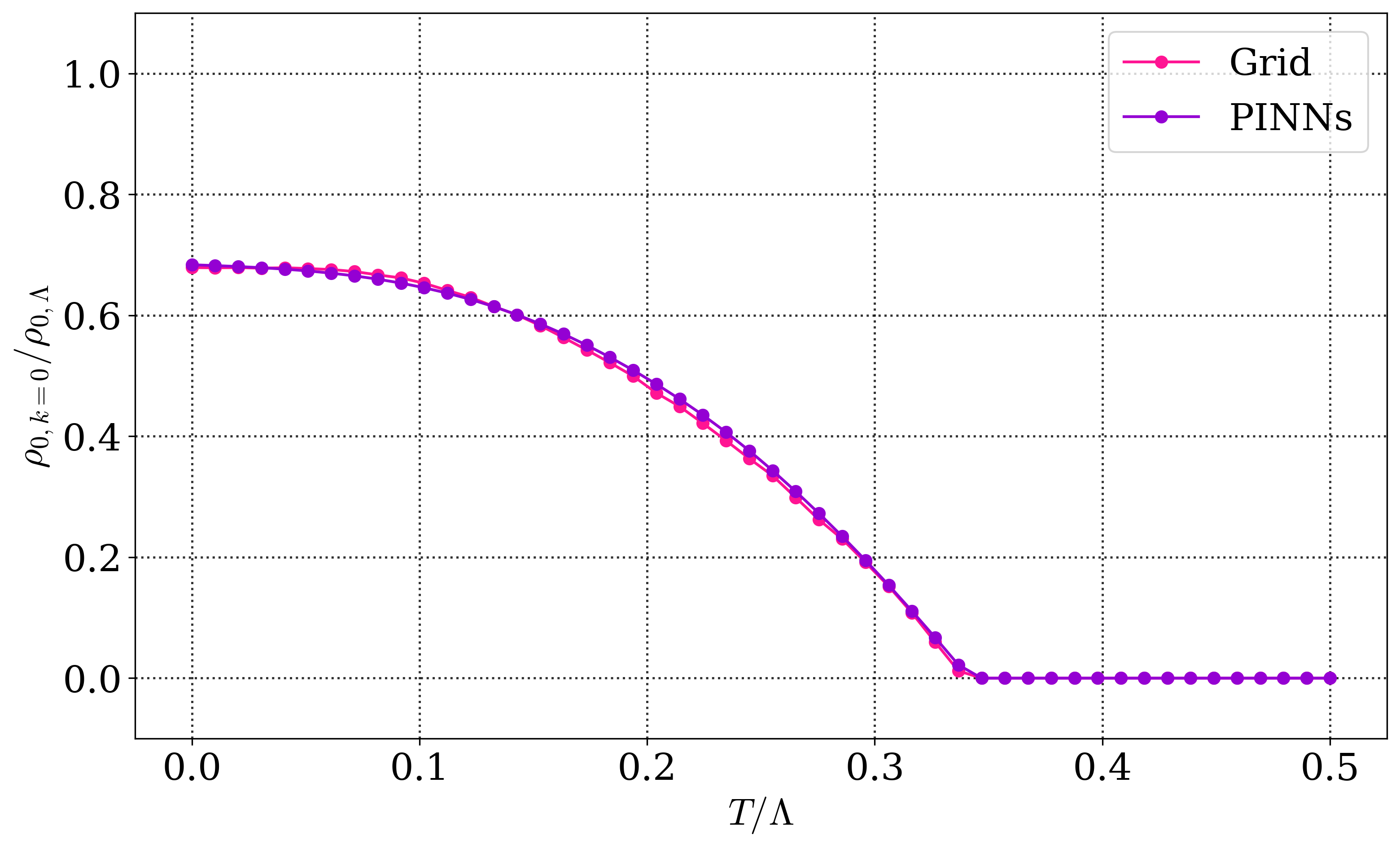}
    		      \caption{Comparing the resulting $\rho_{0,k=0}(T)$ curves of the grid and the PINNs of $\text{HS} = 100$.}
    		      \label{anharm_varyT_HS100_MINPOS}
    	    \end{subfigure}
            \caption{Analyzing the accuracy of learned temperature-dependence of the PINNs by comparing its order parameter--temperature curve with results of the grid method.}
            \label{anharm_varyT_MINPOS_ANAL}
        \end{figure}
    \end{center}
\end{widetext}

Fig. \ref{anharm_varyT_varyHS_MINPOS_DIFF} shows that the PINNs predictions oscillate around the grid results and, as expected, with increasing hidden size the amplitude of the oscillation decreases. Fig. \ref{anharm_varyT_HS100_MINPOS} displays that the $\text{HS} = 100$ result practically reproduces the grid result, proving our claim that the proposed neural network can capture all the $\rho$, $k$ and $T$ dependencies of the effective potential at the same time. This is the main result of this paper.

\section{Conclusions}
In this paper, we presented a novel physics-informed neural networks based approach for solving functional renormalization group flow equations in the local potential approximation. For our investigations, we chose the finite temperature $O(N)$ symmetric field theory. In order to validate results of our neural network, we also employed a traditional grid-based solution of the flow equation. We argued that, for theories with more complex symmetry groups and multiple invariants, traditional grid-based methods can become increasingly impractical due to their inefficient use of the available degrees of freedom and the numerical instabilities introduced by finite-difference approximations. Our results, therefore, suggest that the PINNs approach could provide a promising alternative for studying such theories in future work.

We wish to highlight that, on the contrary to earlier PINNs attempts on solving flow equations in the $O(N)$ model, the approach presented here does not make use of the large-$N$ expansion and it does not rely even partially on any grid based techniques. These methodological assumptions, adopted in earlier works, would have made the resulting approximation schemes unlikely to generalize to more complex theories.

Our method is based on an ansatz that combines small- and large-field expansion schemes as well as incorporating neural networks, yielding a highly accurate approximation across the entire field range. As a result, our approach does not require any discretization, and the degrees of freedom of the neural network are fully decoupled from the domain of the effective potential. In contrast to grid-based methods, this provides considerable flexibility for addressing more complicated flow equations without introducing the numerical instabilities associated with discretization. We presented a detailed description of how we progressively arrived at the final form of the ansatz [see Eq.~(\ref{ansatz_semifinal})], and we also provide the code used for its training~\cite{MyCode}.

We solved the LPA-FRG flow equation using the neural network in two settings. First, we considered a fixed-temperature scenario, in which the model only had to learn the RG flow of $V_k'(\rho)$ at $T=0$. After demonstrating that our ansatz, implemented with a relatively small network (depth $=4$ and hidden size $=40$), could reliably reproduce the RG flow, we extended the training to finite and varying temperatures. In this case, the network had to learn the full dependence of $V_k'(\rho;T)$ on $\rho$, $k$, and $T$. We then investigated in detail how the hidden size of the neural network affects the accuracy of the trained model. We found that increasing the hidden size to $\mathrm{HS}=100$ resulted in a temperature dependence of the order parameter that was practically indistinguishable from that obtained using the grid-based method. These results demonstrate that the neural network not only provides an accurate solution to the flow equation for the effective potential at a fixed temperature, but is also capable of learning its full $T$ dependence.

In future work, the proposed method could be extended to more complicated systems in which the effective potential depends on multiple field variables. Beyond providing a promising fully non-perturbative approach for accessing the scale and temperature dependencies of effective potentials, our method could also be used to search for renormalization group fixed points in the functional sense. Potential applications include scalar theories with $O(N)\times O(M)$ \cite{Yabunaka2025,Sanchez2025} or $SU(N)\times SU(N)$ symmetries \cite{Fejos2024}, as well as an investigation of the applicability of the method to gauge theories. These extensions would provide a natural test of the method in increasingly complex settings and constitute interesting directions for future work.

\section*{Acknowledgments}

The authors are grateful to Barnab\'as Moln\'ar and Nick Horv\'ath for discussions. J.G. was supported by the EKÖP-25 University Research Scholarship Program of the Ministry for Culture and Innovation. G.F. was supported by the Hungarian National Research, Development, and Innovation Fund under Project No. FK142594 and by the Fulbright Program, sponsored by the U.S. Department of State.

\vfill
\bibliography{bibliography}
\end{document}